\documentstyle[aps,epsfig,twocolumn,eqsecnum]{revtex}
\title{Stroboscopic theory of atomic statistics in the micromaser}
\author{R.R.~Puri\cite{byline1}, S.~Arun~Kumar\cite{byline2}, and 
R.K.~Bullough\cite{byline3}}
\address{Department of Mathematics, UMIST, PO Box 88, Manchester M60 1QD,
UK.} 
\date{\today} 
\begin{document}

\twocolumn[\hsize\textwidth\columnwidth\hsize\csname
@twocolumnfalse\endcsname

\maketitle 
\begin{abstract}
We study the statistics of the atoms emerging from the cavity of a
micromaser in a dynamical, discrete-time `stroboscopic' description which
takes into account the measurements made, in general, with imperfect
efficiencies $\eta < 1$, on the states of the outcoming atoms. Inverted
atoms enter stochastically, in general, with a binomial distribution in
discrete time; but we also consider the continuous-time limit of this
input statistics which is Poissonian.  We envisage two alternative
experimental procedures: one of these is to consider a fixed number $N$ of
atoms pumped into the cavity and subsequently leaving it to undergo state
detection; the other is to consider input of the excited atoms and their
subsequent detection and collection in a fixed time $t$.  We consider, in
particular, the steady state behaviors achieved in the two limits,
$N\rightarrow \infty$ and $t\rightarrow \infty$, as well as the approaches
to these two limits. Although these limits are the same for the state of
the cavity field, they are not the same, in general, for the observable
outcoming atom statistics. We evaluate, in particular, Mandel's
$Q$-parameters $Q_{e}$ $(Q_{g})$ for outcoming atoms detected in their
excited states (ground states), for both $N\rightarrow \infty$ and
$t\rightarrow \infty$, as functions of $N_{ex} = RT_{c}$: $R$ is the mean
rate of entry for the incoming atoms and $T_c$ is the cavity damping time.
The behavior of these atomic $Q$-parameters is compared with that
parameter for the cavity field. 
\end{abstract}

\vskip1pc]
\narrowtext

\section{Introduction}
\label{sec:sec1}

The micromaser, in which two-level atoms enter a high-$Q$ cavity one at a
time and interact with a single mode of the cavity field before
subsequently leaving the cavity, provides a valuable means of testing
various aspects of the quantized-field-quantized-atom interaction
\cite{r1,r2,r3,r4,r5,r6,r7,r8,r9,r10,r11,r12,r13,r14}; while the
one-atom-one-mode model of Jaynes and Cummings \cite{r15} used to describe
the strictly one-atom micromaser, as well as the $N$-atom-one-mode model
\cite{r16} for those cases where more than one atom is in the cavity at
the same time, are of fundamental interest in the theory of exactly
solvable quantum models \cite{r11,r17,r18,r19}.  Despite, for example, a
recent operator equations of motions study of the dynamics of these
micromasers \cite{r19}, the body of theoretical investigations has been
largely based on the characteristic properties of the cavity field
averaged over the states of the atoms leaving the cavity. However, because
of the practical difficulties associated with measuring the states of this
cavity field, these theoretical predictions can not be checked
experimentally directly. In the experiments, the measurements are actually
made on the states of the atoms leaving the cavity
\cite{r11,r14,r20,r21,r22}. The properties of the cavity field must then
be inferred from the statistics of the states of the outcoming atoms. It
is, therefore, essential to know the relationship between the field
statistics and this atomic statistics.  This relationship, evidently,
requires the knowledge of not only the field statistics but also that of
the atomic statistics. 

The atomic statistics in the micromaser, namely the outcoming atom
statistics, refers to the probability distribution functions for observing
the atoms in one state or the other after they have left the cavity. These
functions can obviously be found by repeatedly monitoring these states,
either for a fixed number $N$ of the atoms pumped into the cavity and
subsequently leaving it or for the atoms pumped for a fixed duration $t$
of time and subsequently collected in that time. The two descriptions are
evidently equivalent if the time interval between the successive atoms
entering the cavity is a fixed number but not if that time interval is a
random variable. The dynamics of the cavity field can similarly be
described either as a function of the number $N$ of atoms traversing the
cavity or as a function of total passage time $t$. Of particular interest
is the asymptotic, i.e., the steady state, behaviors of the system. The
meaning of these asymptotic limits must, of course, depend upon the chosen
description: the limit corresponds to the limit $N\rightarrow\infty$ if
the dynamics is described in terms of the number of atoms and it
corresponds to the limit $t\rightarrow\infty$ if the dynamics is described
in terms of the collection time. However, it has already been shown
\cite{r8} that the density matrix of the cavity field in both of these
limits is the same. 

The atomic statistics of the micromaser, on the other hand, has only been
studied in terms of the atoms collected in a fixed interval of time and
that too in the `coarse-grained' description of the equations of motion
for the cavity field density operator $\rho_{f}$
\cite{r11,r23,r24,r25,r26,r27,r28,r29}. However, and despite the fact that
coarse-graining appears to involve a coarse averaging over the successive
entries of the atoms \cite{r11,r23}, it has been known \cite{r30} that
when the probability distribution of the intervals between successive
atoms entering the cavity is exponential, i.e., when the pumping mechanism
is Poissonian, coarse-graining which assumes a steady rate $R$ (say) of
entry for the atoms, is precisely equivalent to assuming this strictly
Poissonian pumping of the atoms with the same mean rate $R$. The refs.
\cite{r25,r26,r27,r28,r29} for example adopt this coarse-grained
description from the outset and then collect the outcoming atoms in a
fixed interval of time. It is therefore of interest to see whether or not
these two alternative descriptions of a fixed number $N$ of atoms
collected or of a fixed time $t$ for that collection are asymptotically
equivalent in terms of the outcoming atom statistics rather than the field
statistics for a general pumping statistics which includes Poissonian
pumping only as a special case. Moreover, and despite the fundamental
interest of this problem to ergodic theory, insofar as these two limits
prove to be different (and they do) there is the possibility at the level
of actual experiment of gaining additional information by collecting at
fixed $N$ on the one hand and at fixed $t$ on the other. We shall see that
this additional information can be helpful in determining the actual state
of the cavity field in the micromaser. Thus it is this general question,
of fixed $N$ or of fixed $t$, which is the question addressed in this
paper. 

In this paper we find the probability of the occupation of the two atomic
states (to be called $\vert e\rangle$, excited and $\vert g\rangle$,
ground) for a fixed number $N$ of atoms leaving the cavity and we find the
same probability for the group of atoms collected in a fixed time $t$. We
shall assume the pumping mechanism is binomial. Both regular and
Poissonian pumping are then special cases of this binomial pumping. We
shall evaluate the parameters $Q$ for these two different cases: $Q$ takes
the form of Mandel's `$Q$-parameter' \cite{r31} and is a measure of the
deviation of the variance for the population of a given state from the
value it would have if the distribution of that population were
Poissonian. We shall find that these $Q$-parameters for each of the two
states are generally different in the two asymptotic limits, the limits
$N\rightarrow\infty$ and $t\rightarrow\infty$. We investigate the
dependence of these differences on the parameter $N_{ex}$ where $N_{ex}$
is the average number of atoms entering the cavity in one decay time of
that cavity. The $Q$-parameters $Q_{e}$ for the excited state $\vert
e\rangle$ are found to be different for all $N_{ex}$ in these two
asymptotic limits. However, the $Q$-parameters $Q_{g}$ for the lower state
$\vert g\rangle$ are different in the two asymptotic limits for small
$N_{ex}$ but those differences gradually disappear with an increase in
$N_{ex}$. The value of $N_{ex}$ after which the differences between the
two asymptotic results become insignificant are found to depend upon the
interaction time. At zero temperature it is $Q_{e}$ for a fixed number $N$
of atoms as $N \rightarrow \infty$ which closely follows $Q_{f}$ the
$Q$-parameter for the cavity field: but at finite temperatures this close
relationship changes significantly. These observations concern only the
low lying trapping states: for other equilibrium states of the cavity
field $Q_{e}$ joins substantially in its behavior with the behavior of the
other $Q$'s even at zero temperature. 

The paper is organized as follows:- By following the stroboscopic
description, we derive in section \ref{sec:sec2} expressions for the joint
probability of finding sets of successively entering atoms in a particular
sequence of detected states. We also derive a number of alternative
stroboscopic maps connecting successive states of the cavity field, and
which have taken into account as an ensemble average the different
sequences of atomic state detections. From these in the continuous time
i.e., Poisson limit of the input statistics we derive master equations for
the field density operator for fixed $N$. From this result we can regain
the usual coarse-grained master equation for the field density operator.
In the section \ref{sec:sec3} we use the joint probability found in
section \ref{sec:sec2} to find the joint probability of detecting given
numbers of atoms in each of the two states on the passage of $N$ active
atoms and from it the variances in the number of atoms detected in a given
state. We also find the joint probability of detecting given numbers of
atoms in the two states for a fixed collection time $t$ of the atoms and
from this determine the variances for the atoms collected in the fixed
time $t$. Numerical results comparing the two descriptions are then
presented in the section \ref{sec:sec4}. In an Appendix we outline a
method for calculating $Q_{e}$ and $Q_{g}$ for fixed $N$ and the
corresponding $\tilde Q_{e}$ and $\tilde Q_{g}$ for fixed $t$. 

\section{joint probability for state detection of successive atoms}
\label{sec:sec2}

In this section we derive an expression for the joint probability for
detecting each of a set of successively entering atoms in either one of
their two states after they have left the micromaser cavity. 

We consider the usual micromaser system in which a high-$Q$ cavity is
pumped by Rydberg atoms at a rate $R$ so low that almost always there is
at most one atom at a time in the cavity. We assume that the atoms enter
the cavity prepared in the state $|e\rangle$ which is coupled resonantly
by a cavity mode of frequency $\omega_0$ to a lower energy state
$|g\rangle$ which, for the sake of convenience, we refer to as the ground
state. The interaction between such an atom and the e.m. field as the atom
traverses the cavity is governed by the Jaynes-Cummings Hamiltonian
\cite{r15}
\begin{eqnarray}
H=\hbar\Big[\omega_0 a^\dagger a + \omega_0 S_z + g
\Big(a^\dagger S_-+ S_+ a\Big)\Big],
\label{eq:eq2_1}
\end{eqnarray}
where $(a,a^\dagger)$ are the field annihilation and creation operators
which obey the commutation relation $[a,a^\dagger]=1$ for bosons (they
satisfy the Heisenberg-Weyl algebra \cite{r11,r19}); operators
$S_+=|e\rangle\langle g|,S_-=|g\rangle \langle e|,S_z=(1/2)
[|e\rangle\langle e|-|g\rangle\langle g|]$ are the atomic operators (and
satisfy the su(2) Lie algebra for total spin $S=\frac{1}{2}$
\cite{r11,r19}), and $g$ is the atom-field coupling constant. The cavity
field and the atoms interact also with a heat-bath held at a constant
temperature. This coupling induces the excited atomic state to decay
spontaneously and to the decay of the cavity field. However, the time for
the spontaneous decay between two Rydberg levels, which are the levels of
interest here, is long enough compared with the convenient choices of the
times of flight $t_{int}$ of the atoms through the cavity to enable one to
ignore, to a very good approximation, the effects of the atomic coupling
to the thermal reservoir ($t_{int}$ is the interaction time between the
atoms and the cavity field and is $\sim 35 \mu$ s \cite{r20}). Moreover
the coupling of the cavity field to the heat-bath may also be ignored
during this atom-transit time $t_{int}$ because this time is several
orders of magnitude smaller than the decay time $T_{c}$ of the cavity
field ($T_{c} \sim 0.2$ s \cite{r20}). Together these two acceptable
approximations mean that the dynamical evolution is simply unitary under
the Hamiltonian $H$ given by (\ref{eq:eq2_1}) during the short time
$t_{int}$. Note that the micromaser system has been studied numerically
and in depth during this short atomic transit time $t_{int}$ with the
coupling of the field to the heat-bath included \cite{r10,r11,r12} and it
has also been studied analytically \cite{r11,r32,r33,r34}. Analytical
results for the Jaynes-Cummings Hamiltonian (\ref{eq:eq2_1}) coupled to
the heat-bath have also been given in ref. \cite{r35} in a different
context. All of these studies together confirm the fact that the coupling
to the heat-bath does not play any significant role if $t_{int}$ is short
enough compared with the atomic and field damping times. As noted this is
the case, for example, in the experiments of ref. \cite{r20}---although
\cite{r11} reports some small differences (of the order of percent) from
Meystre's formula \cite{r1} for the probability of finding $n$ photons in
the cavity mode even when $N = 6000$ atoms have passed through the cavity
for parameters within this experimental range. The numerical work
\cite{r11} is essentially exact and the differences of the order of
percent are compatible with the errors involved in the second of the two
approximations itemized above together with the approximation $t_{i+1} -
t_{i} - t_{int} \approx t_{i+1} - t_{i}$ made below; however $N = 6000$
atoms may still not be close enough to equilibrium. 

Still, whatever the exact situation for the short time $t_{int}$, the
coupling of the field with the heat-bath must certainly never be ignored
during the time intervals between the arrivals of two consecutive atoms
into the cavity---for these time intervals are usually significant as
compared with the damping time of the field. If the heat-bath has an
average of $\bar n$ thermal photons at the cavity frequency $\omega_0$ and
if $2 \kappa \equiv T^{-1}_{c}$ is the rate of loss of photons then the
evolution of the density matrix $\rho_f$ of the field due to its
interaction with the thermal reservoir is governed by the usual master
equation taken in the rotating frame (cf. e.g., \cite{r11} and references)
which is
\begin{eqnarray}
{d\rho_{f}\over dt} &=&\kappa[(\bar n+1)(2a\rho_f a^\dagger
-a^\dagger a\rho_f-\rho_f a^\dagger a) \nonumber \\
& & + \bar n(2a^\dagger \rho_f a-aa^\dagger \rho_f-\rho_f aa^\dagger)]
\nonumber\\
& \equiv & L\rho_{f}.
\label{eq:eq2_2}
\end{eqnarray}
Note that $L$ is norm preserving i.e., $\frac{d}{dt} Tr \rho_{f} = Tr L
\rho_{f} = 0$, for the right side is expressible as a sum of commutators
the characteristic form for Markovian dynamics \cite{r36} (actually the
sum in (\ref{eq:eq2_2}) is a sum of double commutators). Now, let
$\rho_f(t_i)$ be the density matrix of the cavity field at the time $t_i$
of the entry of the $i^{th}$ atom into the cavity. Since, in this paper,
the atoms are assumed to enter the cavity in their excited state
$|e\rangle$, it follows that the state of the combined system of the atom
and the field at the time $t_i$ is the outer product of $\rho_f(t_i)$ and
$|e\rangle\langle e|$. The system then evolves, as discussed above, under
the action of the Hamiltonian (\ref{eq:eq2_1}) alone for the time
$t_{int}$---the time of flight of the atom through the cavity. It can then
be shown that under this unitary evolution during $t_{int}$ the density
matrix $\rho(t_i+t_{int})$ of the combined system as the $i^{th}$ atom
leaves the cavity is given by
\begin{eqnarray}
\rho(t_i+t_{int}) & = & \Big[F_e(t_{int})|e\rangle\langle e|+F_g(t_{int})
|g\rangle\langle g| \nonumber \\
& & + F_{eg}(t_{int})|e\rangle\langle
g|+F^\dagger_{eg}(t_{int}) |g\rangle\langle e|\Big] \rho_f(t_i) \nonumber 
\\
\label{eq:eq2_3}
\end{eqnarray}
where
\begin{eqnarray}
F_e(t_{int})\rho_f(t_i) & = & \cos(gt_{int}\sqrt{a^\dagger a +1})
\rho_f(t_i) \nonumber \\
& & \times \cos(gt_{int}\sqrt{a^\dagger a +1}),\nonumber\\
F_g(t_{int})\rho_f(t_i) & = & a^\dagger{1\over 
\sqrt{a^\dagger a +1}}\sin(gt_{int}\sqrt{a^\dagger a +1})\rho_f(t_i)
\nonumber \\
& & \times \sin(gt_{int}\sqrt{a^\dagger a +1}){1\over \sqrt{a^\dagger a
+1}}a, \nonumber\\
F_{eg}(t_{int})\rho_f(t_i)\mbox{}&=& 
i\cos(gt_{int}\sqrt{a^\dagger a +1})\rho_f(t_i) \nonumber \\
& & \times \sin(gt_{int}\sqrt{a^\dagger a +1}){1\over \sqrt{a^\dagger a +
1}}a.
\label{eq:eq2_4}
\end{eqnarray}
Significant quantities are, e.g. \cite{r1,r11,r23,r24,r32,r35}
\begin{eqnarray}
\beta_{m}=\sin^2(gt_{int}\sqrt{m})=1-\alpha_m.
\label{eq:eq2_5}
\end{eqnarray}
The quantity $\beta_{m+1}$ is the {\em a priori} probability that an
excited atom after traversing the cavity containing $m$ photons would exit
it in its ground state and $\alpha_{m+1}$ is the probability of exiting in
its excited state (cf. e.g., \cite{r37}). The expression (\ref{eq:eq2_3})
shows that the actual probability $P(e;t_i)$ $(P(g;t_i))$ that the
$i^{th}$ atom exits the cavity in the excited (ground) state is
\begin{eqnarray}
P(e;t_i) & = & Tr_f[F_e\rho_f(t_i)], \nonumber \\
\qquad \Big( P(g;t_i) & = & Tr_f[F_g\rho_f(t_i)] \Big),
\label{eq:eq2_6}
\end{eqnarray}
where $Tr_f$ denotes the operation of trace over the field. 

After leaving the cavity, the atom passes through a detector which
determines its state. We assume that the time that the atom takes to
arrive at the detector after leaving the cavity is small enough for any
losses that the atom may suffer due to spontaneous emission can be
ignored. However, the probability of detection of the atom in either state
may still not be the same as the probability with which it leaves the
cavity in that state. This is because the detector may not be of unit
efficiency. Following ref. \cite{r21}, for example, we assume that the
detector has an efficiency $\eta_e$ ($\eta_g$) for detecting an atom in
its excited (ground) state. Clearly, the probability $P_d(e;t_i)$
($P_d(g;t_i)$) that the $i^{th}$ atom is detected in the state $|e\rangle$
($|g\rangle$) after leaving the cavity is then given by
\begin{eqnarray}
P_d(\nu_i;t_i)&=&\eta_{\nu_i}P(\nu_i;t_i)\nonumber\\
\mbox{}&\equiv& Tr[F_{\nu_i,d}\rho_f(t_i)], 
\quad \nu_i=e,g
\label{eq:eq2_7}
\end{eqnarray}
where $P(e;t_i)$ ($P(g;t_i)$) is the probability, given by 
(\ref{eq:eq2_6}), that 
the atom leaves the cavity in the state $|e\rangle$ ($|g\rangle$) and
\begin{eqnarray}
F_{\nu,d}=\eta_\nu F_\nu,\qquad \nu=e,g.
\label{eq:eq2_8}
\end{eqnarray}
Here, we have introduced the subscript $d$ on those probabilities and
operators which depend on the detector efficiency and $F_{\nu,d}$, like
$F_{\nu}$, depends on $t_{int}$ as in (\ref{eq:eq2_4}). Since the detector
efficiency is not necessarily unity, there is a finite probability that
the atom passes through the detector without its state being detected. It
is straightforward to see that the probability $P_d(n;t_i)$ that the atom
goes through without any state detection is
\begin{eqnarray}
P_d(n;t_i)=Tr[F_{n,d}\rho_f(t_i)],
\label{eq:eq2_9}
\end{eqnarray}
where
\begin{eqnarray}
F_{n,d}=(1-\eta_e)F_e+(1-\eta_g)F_g,
\label{eq:eq2_10}
\end{eqnarray}
and depends on $t_{int}$. We have thus evaluated the probability of
detection in a given state by (\ref{eq:eq2_7}), or of no state detection
by (\ref{eq:eq2_9}), for an atom as it comes out of the cavity. These
expressions (\ref{eq:eq2_7}) and (\ref{eq:eq2_9}) show that the
probability of detecting an atom in one of its states or that of no
detection is obtained by operating on the density matrix (density
operator) of the field, taken at the time of the entry of that atom, by
the operator $F_{\nu,d}$ (now with $\nu=e,g,n)$ followed by the operation
of trace over the field. Next we use this procedure to find the
probability of a particular outcome of the process of state detection for
the next, i.e. the $(i+1)^{th}$, atom. 

Now, in order to determine the probability of detection or of no detection
of a state of the $(i+1)^{th}$ atom on its exit from the cavity, we need
to know the state of field at the time $t_{i+1}$ of its entry into the
cavity. Since the $i^{th}$ atom is subjected to detection, the state of
the field in the cavity at the time of entry of the $(i+1)^{th}$ atom is
conveniently described by the {\em conditional density operator}
$\rho_f(t_{i+1}|\nu_i;t_i)$ \cite{r38} which characterizes the field at
the time $t_{i+1}$ of entry of the $(i+1)^{th}$ atom into the cavity
under the condition that the $i^{th}$ atom, which entered at $t_{i}$, is
detected at $t_{i} + t_{int}$ in the excited state $\nu_i=e$, is detected
in the ground state $\nu_i=g$, or goes through without any state detection
$\nu_i=n$. By following the procedure outlined following equation
(\ref{eq:eq2_10}) the probability of a particular outcome for the process
of detection on the $(i+1)^{th}$ atom is then found to be
\begin{eqnarray}
P_d(\nu_{i+1};t_{i+1}|\nu_i;t_i)=Tr_f\Big[F_{\nu_{i+1},d}
\rho_f(t_{i+1}|\nu_i;t_i)\Big].
\label{eq:eq2_11}
\end{eqnarray}
Next we must find the expression for the conditional density operator
$\rho_f(t_{i+1}|\nu_i;t_i)$ in terms of the state of the field at the time
$t_i$ of the entry of the $i^{th}$ atom. This expression is found by first
determining the conditional density operator
$\rho_f(t_i+t_{int}|\nu_i;t_i)$ of the field at the time $t_{i} + t_{int}$
of the exit of the $i^{th}$ atom. 

By comparison with (\ref{eq:eq2_3}) it is straightforward to see that the
state of the cavity field at the time of exit of the $i^{th}$ atom
corresponding to each outcome of its detection is (with dependence on
$t_{int}$ included)
\begin{eqnarray}
\rho_f(t_i+t_{int}|\nu_i,t_i) & = & \frac{F_{\nu_i,d}(t_{int})\rho_f(t_i)}
{P_d(\nu_i;t_i)}, \quad \nu_i=e,g,n. \nonumber \\
\label{eq:eq2_12}
\end{eqnarray}
After the exit of the $i^{th}$ atom, the density operator
(\ref{eq:eq2_12}) evolves according to the master equation
(\ref{eq:eq2_2}) for the time interval $t_{i+1}-t_i-t_{int}$ until the
time $t_{i+1}$ when the $(i+1)^{th}$ atom enters. So far in all
quantitative micromaser experiments one chooses as far as is possible
\cite{39} that $t_{i+1} - t_{i} \gg t_{int}$. Hence we let $t_{i+1} -
t_{i} - t_{int} \approx t_{i+1} - t_{i} \equiv \tau_i$. In this way the
state of the cavity field at the time $t_{i+1}$ of the entry of the
$(i+1)^{th}$ atom is given in terms of the state at the earlier time
$t_{i} + t_{int}$ and in terms of the outcome of the measurement on the
atom $i$ by
\begin{eqnarray}
\rho_f(t_{i+1}|\nu_{i};t_i)=D(\tau_i)\rho_f(t_i+t_{int}|\nu_{i};t_i),
\label{eq:eq2_13}
\end{eqnarray}
where
\begin{eqnarray}
D(\tau_i)=\exp(L\tau_i),
\label{eq:eq2_14}
\end{eqnarray}
is the formal solution of (\ref{eq:eq2_2}). On substituting
$\rho_{f}(t_i + t_{int}|\nu_{i};t_i)$ from (\ref{eq:eq2_12}) into
(\ref{eq:eq2_13}) we thus get
\begin{eqnarray}
\rho_f(t_{i+1}|\nu_{i};t_{i}) & = &
\frac{D(\tau_i)F_{\nu_i,d}(t_{int})\rho_f(t_i)}{P_d(\nu_i;t_i)}, \nu_i =
e,g,n; \nonumber \\
\label{eq:eq2_15}
\end{eqnarray}
for the density operator of the field at the time of the entry of the
$(i+1)^{th}$ atom under the condition that the $i^{th}$ atom exits the
cavity in the state $|\nu_i\rangle, \nu_i=e,g$ or goes undetected
($\nu_i=n$). 

The Eq. (\ref{eq:eq2_15}) shows that the state of the field at the time of
entry of the $(i+1)^{th}$ atom corresponding to a particular outcome of
the process of measurement on the $i^{th}$ atom is given by operating on
the density matrix at the time of entry of the $i^{th}$ atom by the
operator $F_{\nu_i,d}$ ($\nu_i=e,g,n$) followed by operation with the
operator $D$ and then normalizing the resulting expression to unit trace. 

The desired expression for the conditional probability
$P_d(\nu_{i+1};t_{i+1}|\nu_i;t_i)$ of atomic detection may now be obtained
by substituting (\ref{eq:eq2_15}) in (\ref{eq:eq2_11}). This expression
therefore reads
\begin{eqnarray}
P_d(\nu_{i+1};t_{i+1}|\nu_i;t_i) & = & \frac{Tr_f\Big[F_{\nu_{i+1},d}
D(\tau_i)F_{\nu_i,d}\rho_f(t_i)\Big]}{P_d(\nu_i;t_i)}. \nonumber \\
\label{eq:eq2_16}
\end{eqnarray}

In what follows we require the joint probability $P(\nu_{i+1};
t_{i+1},\nu_{i};t_i)$ that the outcome of the process of detection on the
$i^{th}$ atom is $\nu_i$ together with the outcome for the $(i+1)^{th}$
atom being $\nu_{i+1}$. This probability can be obtained by using the
relationship
\begin{eqnarray}
P_d(\nu_{i+1};t_{i+1},\nu_{i};t_i) & = & P(\nu_{i+1};t_{i+1}|\nu_{i};t_i)
P_d(\nu_{i};t_i),
\label{eq:eq2_17}
\end{eqnarray}
between the joint and the conditional probabilities. On combining
(\ref{eq:eq2_16}) and (\ref{eq:eq2_17}) we find the desired expression
\begin{eqnarray}
P_d(\nu_{i+1};t_{i+1},\nu_{i};t_i) & = & Tr_f[F_{\nu_{i+1},d} D(\tau_i)
F_{\nu_i,d}\rho_f(t_i)], \nonumber \\
\label{eq:eq2_18}
\end{eqnarray}
for the joint probability of a particular pair of outcomes as a result of
measurement on two successive atoms entering the cavity at times $t_i$ and
$t_{i+1}$. We can now repeat the preceding arguments and show that the
joint probability $P_d(\nu_N;t_N, \ldots ,\nu_2;t_2, \nu_1;t_1)\equiv
P_d(\{\nu_k;t_k\}_N)$ that the outcome of measurements at their exits on
the atoms entering the cavity at times $t_1,t_2, \ldots ,t_N$ is
$\nu_1,\nu_2, \ldots ,\nu_N$ is given by
\begin{eqnarray}
P_d(\{\nu_k;t_k\}_N) & = & Tr_f[F_{\nu_N,d}
D(\tau_{N-1})F_{\nu_{N-1},d} \ldots \nonumber \\
&& \ldots D(\tau_{2}) F_{\nu_2,d} D(\tau_{1}) F_{\nu_1,d} \rho_f(t_1)],
\label{eq:eq2_19}
\end{eqnarray}
where $\rho_f(t_1)$ is the density operator of the field at the time of
entry of first atom into the cavity. 

Similarly, by following the procedure outlined after (\ref{eq:eq2_15}),
the state of the cavity field at the time $t_N$ of the entry of the
$N^{th}$ atom is described by the conditional density operator
\begin{eqnarray}
& &\rho_f(t_N|\{\nu_k;t_k\}_{N-1}) = \nonumber \\
& & ~~~~~~\frac{[D(\tau_{N-1}) F_{\nu_{N-1},d} \ldots D(\tau_{1})
F_{\nu_1,d}\rho_f(t_1)]}{P_d(\{\nu_k;t_k\}_{N-1})}. \nonumber \\
\label{eq:eq2_20}
\end{eqnarray}
Note that the density matrix $\rho_f(t_N)$ characterizing the state of the
cavity field without any condition on the state of the exiting atoms must
be related to the conditional state by the relation
\begin{eqnarray}
\rho_f(t_N) & = &\sum_{\{\nu_k=e,g,n\}}
\rho_f(t_N|\{\nu_k;t_k\}_{N-1}) P_d(\{\nu_k;t_k\}_{N-1}), \nonumber \\
\label{eq:eq2_21}
\end{eqnarray}
where the summation denotes the sum over all the possible outcomes. On
substituting (\ref{eq:eq2_20}) in (\ref{eq:eq2_21}) we get
\begin{eqnarray}
\rho_f(t_N) & = & \sum_{\{\nu_k=e,g,n\}}\Big[D(\tau_{N-1}) F_{\nu_{N-1},d}
\ldots \nonumber \\
& & \ldots D(\tau_2) F_{\nu_2,d} D(\tau_1) F_{\nu_1,d}
\rho_f(t_1)\Big].
\nonumber \\
\label{eq:eq2_22}
\end{eqnarray}
The summation on each of the $\nu_k$ for one label $k$ can now be carried
out by noting that for each $k$
\begin{eqnarray}
\sum_{\nu_k=e,g,n}F_{\nu_k,d}\equiv F_{e,d}+F_{g,d}+F_{n,d}
=F_e+F_g,
\label{eq:eq2_23}
\end{eqnarray}
where the equality follows from the use of (\ref{eq:eq2_7}) and
(\ref{eq:eq2_10}). Note that this step at (\ref{eq:eq2_23}) {\em neatly
eliminates} the detector efficiencies $\eta_{e}$, $\eta_{g}$ from
(\ref{eq:eq2_22}). For a first investigation let us now assume that the
atoms arrive with a regular spacing $\tau_p$, i.e., $\tau_i$ is $\tau_p$
for all $i$. The Eq. (\ref{eq:eq2_22}) can then be seen to be exactly
\begin{eqnarray}
\rho_f(t_N)=[D(\tau_p)F_0]^{N-1}\rho_f(t_1),
\label{eq:eq2_24}
\end{eqnarray}
where
\begin{eqnarray}
F_0=F_e+F_g.
\label{eq:eq2_25}
\end{eqnarray}
Note from (\ref{eq:eq2_3}) that $F_0\rho(t_i)$ is the density operator of
the field at the time of exit of the $i^{th}$ atom if the state of this
outcoming atom is left undetermined. Moreover the Eq. (\ref{eq:eq2_24})
shows that, as is required by (\ref{eq:eq2_22}), {\em the ensemble
averaged field does not depend in any way upon the efficiencies of the
detectors}. 

The Eq. (\ref{eq:eq2_24}) is evidently equivalent to the stroboscopic
equation or `map'
\begin{eqnarray}
\rho_f(t_{i+1})=D(\tau_p)F_0\rho_f(t_i),
\label{eq:eq2_26}
\end{eqnarray}
for the regular input as it was used in e.g. ref. \cite{r13}. The route by
which (\ref{eq:eq2_26}) is arrived at here is, of course, not at all the
one that is usually followed to reach this expression. The usual
derivation (e.g. \cite{r13}) does not make any reference, from the outset,
to the processes of atomic state measurement. The derivation given above
specifically demonstrates the consistency of that approach with that
developed here now taking proper account of the process of atomic
measurement. 

The derivation of (\ref{eq:eq2_26}) given above shows actually that, for
regular inputs at least, all effects of the atomic state detection
processes described by the efficiencies $\eta_{i}$ vanish under the
ensemble average. Indeed because of (\ref{eq:eq2_23}) these efficiencies
vanish through the ensemble average (\ref{eq:eq2_22}) for
$\rho_{f}(t_{N})$ for {\em any} statistics of the sequence $\tau_{i}$ of
successive atomic inputs. For the cavity field therefore, effects of state
detection have vanished naturally from the ensemble averaged theory.
Actual effects of the detection process via the atoms on the state of the
cavity field were envisaged already in \cite{r1} and were investigated via
numerical methods in \cite{r21}; one of the present authors also
investigated such effects numerically \cite{r11,r40} by however studying
only single realizations of the atomic inputs---where substantial effects
can be discerned \cite{r40}. Within coarse-grained theory (that is,
equivalently, for continuous time Poisson inputs---see below) such effects
of the atomic detection processes on the state of the cavity field were
also investigated in \cite{r25}. Also within coarse-grained theory, by
introducing a remarkable {\em non-linear} master equation taking account
of the outcomes of the atomic measurements the ref. \cite{r26} shows how
these measurement process effects are ultimately eliminated via the
ensemble averaging within this theory. Our step at (\ref{eq:eq2_23}) on
the ensemble averaged, and usual linear, density operator for the field,
shows that quite generally for discrete time entries of the atoms it is
not possible to detect via the atoms any effect of the measurements on the
atoms on the state of the cavity field. This last statement follows from
the fact that we show in section \ref{sec:sec3} how the variances for the
observed atom statistics depend only on the ensemble averaged field
density operator $\rho_{f}(t_{N})$. Of course this expression
$\rho_{f}(t_{N})$, Eq. (\ref{eq:eq2_22}), must itself be ensemble averaged
over the individual realizations of the sequence $\tau_{i}$, of the atomic
inputs. In the derivation of (\ref{eq:eq2_24}) we took the $\tau_{i}$ to
be constants, $\tau_{i} = \tau_{p}$ for all $i$. We also assumed even in
(\ref{eq:eq2_22}) that the interaction time $t_{int}$ was a constant.
However, in actual experimental situations, the time difference $\tau_i$
between the arrival of successive atoms will not be the same for all the
atoms. And even the time of interaction $t_{int}$ of the atoms may vary
because different atoms may travel with different speeds. In fact, in
practice, the times $\tau_i$ and $t_{int}$ must both be taken to be random
variables. 

In this paper we shall consider only the variations in the pumping time
by following the widely used model \cite{r1,r3} of pumping time
fluctuations. According to this model, atoms leave the source oven at
regular intervals $T$ (say) and are then prepared in a given excited state
$\vert e\rangle$ with a certain probability $p$ prior to their entering
the cavity in that excited state. Thus atoms entering with regular
spacings $T$ have only the probability $p$ of being in $\vert e\rangle$
and only atoms arriving in $\vert e\rangle$ are effective in determining
the dynamics of the micromaser. Even so one may still describe this
dynamics in terms of the number $K$ of all the atoms or equivalently in
terms of the discrete time $t_{K} = KT$ that all these $K$ atoms take to
arrive in the cavity. Alternatively the dynamics may be described in terms
of the number $N$ of atoms entering the cavity specifically in their
active excited states $|e\rangle$. 

First we find the expression for the joint probability $P_d(\{
\nu_k;t_k\}_N,t_K)$ that $N$ atoms enter the cavity in their excited
states $|e\rangle$ in time $t_K$ and that the outcome of the measurement
on those atoms after they leave the cavity is $\nu_1,\nu_2, \ldots 
,\nu_N$. To find this probability let us assume that the state of the
cavity field at the time $t=0$ is $\rho_f(0)$. Now, assume that the atoms
numbered $k_i$, $i = 1, 2, \ldots ,N$ are excited to the respective state
$|e\rangle$. The probability for this event is obtained by using the fact
that the probability of $m$ failures followed by a success in a binomial
process in which the probability of a success is $p$ is given by
$p(1-p)^m$. The time interval between the arrival of active atoms numbered
$i-1$ and $i$ ($i=1, 2, \ldots ,N$) is therefore given by $(k_i -
k_{i-1})T$ with $k_0 \equiv 1$. The probability $P_d(\{\nu_k;
t_k\}_N,t_K)$ can be obtained from (\ref{eq:eq2_19}) (i) by replacing the
damping operator $D(\tau_{i})$ between the atoms numbered $i$ and $i-1$ by
$D((k_i-k_{i-1})T)$, (ii) by noting that the field density operator
$\rho_f(t_1)$ at the time of the entry of the first active atom is
$D((k_1-1)T)\rho_f(0)$ and (iii) by performing the summation over all
$k_i$ to account for all possible realizations of $\{k_i\}$. It then
follows that
\begin{eqnarray}
& &P_d(\{\nu_k;t_k\}_N,t_K) = p^N(1-p)^{K-N} \sum_{k_N=N}^{K}
\sum_{k_{N-1} = N-1}^{k_N-1} \ldots \nonumber \\
& & \ldots \sum_{k_1=1}^{k_2-1}
Tr_f\Big[D((K-k_{N}+1)T)F_{\nu_N,d} D((k_N-k_{N-1})T)\nonumber \\
& & \times F_{\nu_{N-1},d} \ldots D((k_2-k_1)T)F_{\nu_1,d}D((k_1-1)T)\rho_f(0)\Big],
\nonumber \\
\label{eq:eq2_27}
\end{eqnarray}
where, for later manipulative convenience, we have multiplied on the left
by $D((K - k_{N} + 1)T)$ the factor appearing under the trace. This does
not change the results because of the fact that $L$ is norm preserving and
$Tr(L\rho) = 0$ so that
\begin{eqnarray}
Tr(D\rho)=Tr(\rho),
\label{eq:eq2_28}
\end{eqnarray}
for any $\rho$.

The evaluation of (\ref{eq:eq2_27}) is facilitated by defining the
(probability) generating function
\begin{eqnarray}
f(N,x)=\sum_{K=N}^{\infty}x^{K} P_d(\{\nu_k;t_k\}_N,t_K),
\label{eq:eq2_29}
\end{eqnarray}
so that 
\begin{eqnarray}
P_d(\{\nu_k;t_k\}_N,t_K)={1\over K!}{d^K\over dx^K}f(N,x)|_{x=0}.
\label{eq:eq2_30}
\end{eqnarray}
On using the expression (\ref{eq:eq2_27}) for
$P_{d}(\{\nu_{k};t_{k}\}_{N},t_{K})$ in (\ref{eq:eq2_29}) and on carrying
out the summations over the $\{k_i\}$ and $K$ we get
\begin{eqnarray}
f(N,x) & = & (px)^NTr_f[D_p(x)\exp(LT)F_{\nu_N,d}D_p(x) \exp(LT) \nonumber
\\
& & \times F_{\nu_{N-1},d} \ldots D_p(x) \exp(LT)F_{\nu_1,d} D_p(x)
\rho_f(0)], \nonumber \\
\label{eq:eq2_31}
\end{eqnarray}
where
\begin{eqnarray}
D_p(x)={1\over 1-x(1-p)\exp(LT)}.
\label{eq:eq2_32}
\end{eqnarray}
Note that the joint probability $P_d(\{\nu_k;t_k\}_N)$ for the outcomes
$\{\nu_k\}$ on the passage of $N$ active atoms irrespective of the time
$t_K$, obtained from $P_d(\{\nu_k;t_k\}_N,t_K)$ by performing the
summation over $K$, is given by
\begin{eqnarray}
P_d(\{\nu_k;t_k\}_N) & = & \sum_{K=N}^{\infty}
P_d(\{\nu_k;t_k\}_N,t_K)=f(N,1),
\label{eq:eq2_33}
\end{eqnarray}
where the last equality is by virtue of (\ref{eq:eq2_29}). The joint
probability $P_d(\{\nu_k;t_k\},t_K)$ for the outcomes $\{\nu_k\}$ in time
$t_K$ irrespective of the number of active atoms is obtained, on the other
hand, by performing the summation over $N$ in $P_d(\{\nu_k;t_k
\}_N,t_{K})$, i.e.,
\begin{eqnarray}
P_d(\{\nu_k;t_k\},t_K)=\sum_{N=0}^{K}P_d(\{\nu_k;t_k\}_N,t_K).
\label{eq:eq2_34}
\end{eqnarray}
We will use (\ref{eq:eq2_33}) and (\ref{eq:eq2_34}) in the next section to
find the probability that a given number of atoms are detected in a given
state after the passage of a fixed number $N$ of active atoms irrespective
of time or that for a fixed collection time $t_{K}$. Here we demonstrate
the consistency of the approach developed so far by deriving the equation
of evolution for the cavity field averaged over all possible outcomes of
detection. 

The cavity field averaged over all possible outcomes of detection at the
time of entry of the $K^{th}$ atom when $N$ atoms have entered the cavity
in their active excited states $\vert e\rangle$ follows from
(\ref{eq:eq2_22}) and the considerations outlined above and can be seen to
be given by
\begin{eqnarray}
& &\rho_f(N,t_K) = p^N(1-p)^{K-N} \sum_{\{\nu_i=e,g,n\}}
\sum_{k_n=1}^{K} \sum_{k_{n-1}=1}^{k_n-1} \ldots \nonumber \\
& & \ldots \sum_{k_1=1}^{k_2-1} 
\Big[D(T(K+1-k_N))F_{\nu_N,d} D(T(k_N-k_{N-1})) \nonumber \\
& & \times F_{\nu_{N-1},d} \ldots D(T(k_2-k_1)) F_{\nu_1,d} D(T(k_1-1))
\rho_f(0)\Big]. \nonumber \\
\label{eq:eq2_35}
\end{eqnarray}
On carrying out the summation over $\nu_i$ in (\ref{eq:eq2_35}) by using
(\ref{eq:eq2_23}) we get
\begin{eqnarray}
& &\rho_f(N,t_K) = p^N(1-p)^{K-N} \sum_{k_n=1}^{K}
\sum_{k_{n-1}=1}^{k_n-1} \ldots \nonumber \\
& & \ldots \sum_{k_1=1}^{k_2-1} \Big[D(T(K+1-k_N)) F_0
D(T(k_N-k_{N-1})) \nonumber \\
& & \times F_0 \ldots D(T(k_2-k_1))F_0D(T(k_1-1)) \rho_f(0)\Big],
\nonumber \\
\label{eq:eq2_36}
\end{eqnarray}
where $F_0$ is again as defined in (\ref{eq:eq2_25}) and the averaging
over the $\nu_{i}$ has again eliminated the effects of the atomic
detection processes. From (\ref{eq:eq2_36}) it is now straightforward to
derive the recurrence relation
\begin{eqnarray}
\rho_f(N,t_{K+1}) & = & \exp(LT)\Big[(1-p)\rho_f(N,t_K) \nonumber \\
& & + p F_0\rho_f(N-1,t_K)\Big],
\label{eq:eq2_37}
\end{eqnarray}
which in principle determines the ensemble averaged cavity field
$\rho_{f}(N,t_{K})$ at the time $t_{K}$ under the condition that in that
time $N$ atoms have entered the cavity in their active excited states
$|e\rangle$. Note once more that this ensemble averaged cavity field,
relation (\ref{eq:eq2_37}) is independent of the detector efficiencies. It
can be seen to reduce to (\ref{eq:eq2_26}) for regular inputs by setting
$p = 1$ and $T = \tau_{p}$. 

The relation (\ref{eq:eq2_37}) can now be used to determine the field
density matrix at the time $t_K$ or at the time of entry of the $N^{th}$
excited atom. The field density matrix $\tilde\rho_f(t_K)$ at the time
$t_K$ is given by $\rho_f(N,t_K)$, summed over all $N$, whereas the field
density matrix $\rho_f(N)$ after the passage of a fixed number $N$ of
active atoms is given by $\rho_f(N,t_K)$, summed over all $K$. Now, on
performing the summation over $N$ in (\ref{eq:eq2_37}) we arrive exactly
at the familiar map
\begin{eqnarray}
\tilde\rho_f(t_{K+1})=\exp(LT)\Big[(1-p)+pF_0\Big]\tilde\rho_f(t_K),
\label{eq:eq2_38}
\end{eqnarray}
already derived in ref. \cite{r8}. The derivation of (\ref{eq:eq2_38}) in
\cite{r8} was however quite different from that just given now; in
\cite{r8} no account whatsoever was taken of the sets of measurement on
the atoms; here these sets of measurement are fundamental to the whole
analysis eventhough they play no role in the final result. In the same way
the summation over $K$ on (\ref{eq:eq2_37}) together with a readjustment
of the terms leads to the map
\begin{eqnarray}
\rho_f(N+1)={p\exp(LT)F_0\over 1-(1-p)\exp(LT)}\rho_f(N).
\label{eq:eq2_39}
\end{eqnarray}
The relation (\ref{eq:eq2_39}) can also be derived from the map
(\ref{eq:eq2_26}) by assuming that the pumping time $\tau_p$ in it is a
random variable corresponding to the binomial process and on performing
the average over all of its possible realizations. It is this procedure
which was followed in ref. \cite{r1} in order to derive the equation for
the case of strictly Poisson pumping. 

Of particular interest is the steady state solution of the maps
(\ref{eq:eq2_38}) and (\ref{eq:eq2_39}). The steady state of the map
(\ref{eq:eq2_38}) ((\ref{eq:eq2_39})) is the state reached in the
asymptotic limit $t_K\rightarrow\infty$ ($N\rightarrow\infty$) and is such
that $\tilde\rho_f(t_{K+1}) = \tilde\rho_f(t_K)$ ($\rho_f(N+1) =
\rho_f(N)$). Hence the steady state $\rho_{ss}$ of (\ref{eq:eq2_38}) is
the solution of the operator equation
\begin{eqnarray}
\Big[1-\exp(LT)\Big((1-p)+pF_0\Big)\Big]\rho_{ss}=0.
\label{eq:eq2_40}
\end{eqnarray}
However it is easy to see that the steady state of the map
(\ref{eq:eq2_39}) also obeys (\ref{eq:eq2_40}). {\em Hence the steady
state density matrix of the cavity field is the same whether the dynamics
is described in terms of time or in terms of the number of active atoms
traversing the cavity.} The exact solution of (\ref{eq:eq2_40}) is known
only for the case of Poisson pumping discussed next. 

A widely studied and experimentally important case of random pumping is
Poisson pumping. This corresponds to the limit $p\rightarrow 0$,
$T\rightarrow 0$ of binomial pumping such that $p/T\rightarrow R$ where
the constant $R$ is the average rate of pumping of the active atoms. In
this limit, the atoms enter the cavity continuously. In other words, in
the Poisson limit, the discrete time $t_K$ becomes a continuous variable
$t$. Hence one can write $\rho_f(N,t_{K+1}) - \rho_f(N,t_K) =
T(d/dt)\rho_f(N,t)$. The recurrence relation (\ref{eq:eq2_37}) then
reduces to the equation
\begin{eqnarray}
{d\over dt}\rho_f(N,t) & = & \Big[L\rho_f(N,t) 
+ R(F_0-1)\rho_f(N-1,t)\Big], \nonumber \\
\label{eq:eq2_41}
\end{eqnarray}
coupling $\rho_f(N,t)$ and $\rho_f(N-1,t)$ and which on iteration
determines the evolution of the field at time $t$ under the condition that
$N$ active atoms enter the cavity in that time. The equation for the field
density matrix $\tilde\rho_f(t)$ at time $t$ irrespective of the number of
active atoms can be derived from (\ref{eq:eq2_41}) by performing the
summation over $N$. This leads to the exact master equation
\begin{eqnarray}
{d\over dt}\tilde\rho_f(t)=\Big[L+R(F_0-1)\Big]
\tilde\rho_f(t).
\label{eq:eq2_42}
\end{eqnarray}
This master equation is the same as the one that is obtained in the
coarse-grained approximation without making any reference to the pumping
statistics (see for example \cite{r26} or \cite{r11}, \cite{r23}---where
coarse-grained averages are actually taken). The Eq. (\ref{eq:eq2_42}) can
also be derived by taking the Poisson limit of (\ref{eq:eq2_38}). Finally,
the density matrix of the field after the passage of a fixed number $N$ of
active atoms, taken irrespective of the time for the case of Poissonian
pumping, is found from (\ref{eq:eq2_39}) to be given by
\begin{eqnarray}
\rho_f(N+1)={1\over 1-L/R}F_0\rho_f(N).
\label{eq:eq2_43}
\end{eqnarray}
This equation is the same as the one derived in ref. \cite{r1} for the
Poisson pumping process achieved by starting from (\ref{eq:eq2_26}) and
assuming the pumping times $\tau_p$ to be random variables corresponding
to a Poisson process and then performing the average over all realizations
of this pumping time. 

Now cf. Eq. (\ref{eq:eq2_27}) and observe that we have there the exact
stroboscopic expression for the probability that $N$ atoms in their active
excited state are pumped into the cavity in a given time $t_K$ and that
the process of state detection after these atoms exit the cavity gives a
particular sequence of outcomes. We can use this expression to study
various aspects of the {\it atomic} statistics including the correlations
between the atoms exiting the cavity at different times. We shall study
these correlations in a later paper. Here we restrict attention to the
study of the probability of detecting a given number of atoms in a given
state. 

\section{Distribution Function for atomic state populations}
\label{sec:sec3}

In the last section we determined at Eq. (\ref{eq:eq2_27}) the joint
probability $P_d(\{\nu_k;t_k,\}_N,t_K)$ for the outcomes $\{\nu_k=e,g,n\}$
as a result of measurement on $N$ excited atoms passing through the cavity
in time $t_K$. Here we use that joint probability to determine the number
distribution $w(N_e,N_g,N)$ which is the probability of detecting $N_e$
atoms in the excited state and $N_g$ atoms in the ground state on the
passage of $N$ active atoms and also the number distribution $\tilde
w(N_e,N_g,t_K)$ for a fixed time interval $t_K$ of observation. We use
these expressions to find the variances in the number of atoms exiting the
cavity in one state or the other. 

Recall that the probability $P_d(\{\nu_k;t_k\}_N)$ is the probability that
the outcome of state measurement on $N$ atoms passing through the cavity
in an arbitrary time is the set $\{\nu_k\}$ ($\nu_k=e,g,n$). This
probability is given by (\ref{eq:eq2_33}). Now, the probability
$w(N_e,N_g,N)$ that $N_e$ atoms are detected in the state $|e\rangle$ and
$N_g$ in the state $|g\rangle$ on the passage of $N$ excited atoms is
clearly the sum of all those joint probabilities $P_d(\{\nu_k;t_k\}_N)$
($\nu_k=e,g,n$) in which $N_e$ of the $\nu$'s correspond to the excited
state and $N_g$ of the $\nu$'s correspond to the ground state. We know
from (\ref{eq:eq2_33}) that $P_d(\{\nu_k;t_k\}_N)=f(N,1)$ where $f(N,x)$
is given by (\ref{eq:eq2_31}). It is then straightforward to see that
$w(N_e,N_g,N)$ is the coefficient of $y^{N_e}z^{N_g}$ in the expansion of
$Tr\Big\{\Big[D_p(1)p\exp(LT)\Big(F_{n,d} + yF_{e,d} +
zF_{g,d}\Big)\Big]^N D_p(1)\rho_f(0)\Big\}$ where $D_p(x)$ is given by
(\ref{eq:eq2_32}). In other words
\begin{eqnarray}
w(N_e,N_g,N) & = & {1\over N_e!N_g!}{d^{N_e+N_g}\over dy^{N_e}dz^{N_g}}
Tr\Big\{\Big[\Lambda_p \nonumber \\
& & \times \Big(F_{n,d}+yF_{e,d}+zF_{g,d}\Big)
\Big]^N\rho_f(t_1)\Big\}\Big|_{y=0}^{z=0}, \nonumber \\
\label{eq:eq3_1}
\end{eqnarray}
where $\rho_f(t_1)=D_p(1)\rho_f(0)$ is the density operator of the field
at the time of entry of the first excited atom,
\begin{eqnarray}
\Lambda_p \equiv D_p(1)p\exp(Lp/R),
\label{eq:eq3_2}
\end{eqnarray}
and we have substituted $T=p/R$ where $R$ is the average rate at which the
atoms are excited. On substituting for $F_{n,d}$ from (\ref{eq:eq2_10})
and on changing the variables $y$ and $z$ to $1-y$ and $1-z$ respectively,
one finds (\ref{eq:eq3_1}) becomes
\begin{eqnarray}
w(N_e,N_g,N) & = & {(-)^{(N_e+N_g)}\over N_e!N_g!}
{d^{N_e+N_g}\over dy^{N_e}dz^{N_g}} Tr\Big\{\Big[\Lambda_p \nonumber \\
& & \times \Big(F_0-y\eta_e F_e-z\eta_g F_g\Big)\Big]^N
\rho_f(t_1)\Big\}\Big|_{y=1}^{z=1}. \nonumber \\
\label{eq:eq3_3}
\end{eqnarray}
The distribution for the number of atoms detected in the ground state
alone, obtained by summing (\ref{eq:eq3_3}) over $N_e$, thus reads
\begin{eqnarray}
w_g(N_g,N) & = & {(-)^{N_g}\over N_g!}
{d^{N_g}\over dz^{N_g}} Tr\Big\{\Big[\Lambda_p\Big(F_0-z\eta_g 
F_g\Big)\Big]^N \nonumber \\
& &\times \rho_f(t_1)\Big\}\Big|_{z=1},
\label{eq:eq3_4}
\end{eqnarray}
together with a similar expression for $w_e(N_e,N)$ in which the
quantities with suffix $g$ are replaced by the corresponding ones with
suffix $e$. Note that $N_e$ and $N_g$ are not independent variables, so
their distributions are related. For the case of unit detection efficiency
for each of the two states, $N=N_e+N_g$ which implies
$w_e(N_e,N)=w_g(N-N_e,N)$. On using the easily verifiable relation
\begin{eqnarray}
& & {1\over k!}{d^k\over dx^k}\Big(A+xB\Big)^N\Big|_{x=0} = \nonumber \\
& & {1\over (N-k)!}{d^{N-k}\over dx^{N-k}}\Big(xA+B\Big)^N\Big|_{x=0},
\label{eq:eq3_5}
\end{eqnarray}
it can then be verified that $w_g(N_g,N)$ given by (\ref{eq:eq3_4}) and
the corresponding expression for $w_e(N_e,N)$ indeed satisfy the relation
$w_e(N_e,N)=w_g(N-N_e,N)$. 

Next we evaluate the atomic population distributions for the case when the
observation is made for a fixed time $t_K$ rather than for a fixed number
of atoms. The joint probability $P_d(\{\nu_k;t_k\},t_K)$ for the outcomes
of detection in that case is determined by working from (\ref{eq:eq2_34})
rather than (\ref{eq:eq2_33}). The probability $\tilde w(N_e,N_g,t_K)$
that $N_e$ atoms are detected in the state $|e\rangle$ and $N_g$ in the
state $|g\rangle$ in time $t_K$ is clearly the sum over all those joint
probabilities $P_d(\{\nu_k;t_k\},K)$ ($\nu_k=e,g,n$) in which $N_e$ of the
$\nu$'s correspond to the excited state and $N_g$ of the $\nu$'s
correspond to the ground state. With $P_d(\{\nu_k;t_k\},t_K)$ given by
(\ref{eq:eq2_34}) taken with (\ref{eq:eq2_30}) and (\ref{eq:eq2_31}) it
follows that
\begin{eqnarray}
& & \tilde w(N_e,N_g,t_K) = {1\over N_e!N_g!K!}{d^{N_e+N_g+K}
\over dy^{N_e}dz^{N_g}dx^K}\sum_{N} (px)^N \nonumber \\
& & \times Tr\Big\{\Big[D_p \exp(Lp/R)
\Big(F_0-(1-y) \eta_eF_e-(1-z) \eta_gF_g\Big)\Big]^N \nonumber \\
& & \times D_p\rho_f(0) \Big\}\Big|_{x=y=z=0}.
\label{eq:eq3_6}
\end{eqnarray}
Now, on performing the summation over $N$ and on rearranging the terms,
one finds that the Eq. (\ref{eq:eq3_6}) yields
\begin{eqnarray}
& & \tilde w(N_e,N_g,t_K) = \frac{1}{N_e!N_g!K!}
\frac{d^{N_e+N_g+K}}{dy^{N_e}dz^{N_g} dx^{K}} \nonumber \\
& & \times Tr\Big\{\Big[1-x\exp(Lp/R) \Big(1-p+p(F_0-(1-y) 
\eta_eF_e \nonumber \\
& & ~~~~~~-(1-z) \eta_gF_g)\Big)\Big]^{-1} \rho_f(0)\Big\}\Big|_{x=y=z=0}. 
\label{eq:eq3_7}
\end{eqnarray}
The differentiation with respect to $x$ can now be performed trivially
to get
\begin{eqnarray}
\tilde w(N_e,N_g,t) & = & {(-)^{N_g+N_e}\over N_e!N_g!}
{d^{N_e+N_g}\over dy^{N_e}dz^{N_g}}Tr\Big\{\Big[\tilde
\Lambda_p \nonumber \\
& & \Big(\tilde F_0-p(y\eta_eF_e+z\eta_gF_g)\Big)
\Big]^{Rt/p}\rho_f(0)\Big\}\Big|_{y=1}^{z=1}, \nonumber \\
\label{eq:eq3_8}
\end{eqnarray}
where
\begin{eqnarray}
\tilde\Lambda_p \equiv \exp(Lp/R), \qquad \tilde F_0 \equiv 1 - p + p
F_{0}.
\label{eq:eq3_9}
\end{eqnarray}
We have replaced $t_K$ by $t$ and substituted $K=t_K/T=Rt/p$. The
probability of detecting $N_g$ atoms in the ground state irrespective of
the number of atoms in the upper state in time $t$, found by performing
the summation over $N_e$ in (\ref{eq:eq3_8}), then reads
\begin{eqnarray}
\tilde w_g(N_g,t) & = & {(-)^N_g\over N_g!}{d^{N_g}\over dz^{N_g}}
Tr\Big[\tilde\Lambda\Big(\tilde F_0-zp\eta_gF_g\Big)^{Rt/p} \nonumber \\
& & \times \rho_f(0)\Big]\Big|_{z = 1},
\label{eq:eq3_10}
\end{eqnarray}
in which $\tilde\Lambda \equiv \tilde\Lambda_{p}$ and there is a similar
expression for $\tilde w_e(N_e,t)$ obtained by replacing the quantities
with the suffix $g$ by the corresponding ones with the suffix $e$. 

In the continuous time Poisson limit $p\rightarrow 0$ we recover from
(\ref{eq:eq3_8}) for fixed $t$ the expression
\begin{eqnarray}
\tilde w(N_e,N_g,t) & = & {(-)^{N_g+N_e}\over N_e!N_g!}{d^{N_e+N_g}
\over dy^{N_e}dz^{N_g}} Tr\Big\{\exp\Big[\Big(L + R(F_0-1) \nonumber \\
& & - y R \eta_{e} F_{e} - z R \eta_{g} F_{g} \Big)t\Big] \rho_f(0)
\Big\} \Big|_{y=1}^{z=1}, \nonumber \\
\label{eq:eq3_11}
\end{eqnarray}
derived in ref. \cite{r26} by their very different methods. The
expressions for $\tilde w_g(N_g,t)$ and $\tilde w_e(N_e,t)$ given in
\cite{r26} are similarly recovered in the Poisson limit of
(\ref{eq:eq3_10}) and from the Poisson limit of the corresponding
expression for $\tilde w_e(N_e,t)$. 

Next we use the number distributions to find the variances in the numbers
of atoms detected in one state or the other. It is straightforward to see
from the expressions for $w_g(N_g,N)$ and $\tilde w_e(N_e,t)$ that at
fixed $N$ for the ground states $\vert g\rangle$
\begin{eqnarray}
\langle N_g\rangle_N & = & -{d\over dz}
Tr\Big\{\Big[\Lambda\Big(F_0-z\eta_g F_g\Big)\Big]^N\Big\}
\rho_f(t_1)\Big|_{z=1}, \nonumber \\
\label{eq:eq3_12}
\end{eqnarray}
and that
\begin{eqnarray}
\langle N_g(N_g-1)\rangle_N & = & {d^2\over dz^2}
Tr\Big\{\Big[\Lambda\Big(F_0-z\eta_g F_g\Big)\Big]^N 
\rho_f(t_1)\Big\}\Big|_{z=1}, \nonumber \\
\label{eq:eq3_13}
\end{eqnarray}
in which $\Lambda \equiv \Lambda_{p}$ while there are similar expressions
for the excited states $\vert e\rangle$ obtained by replacing the
quantities with the index $g$ by the corresponding ones with the index
$e$. The corresponding expressions for a fixed observation time $t$ are
given by
\begin{eqnarray}
\langle N_g\rangle_t & = & -{d\over dz}
Tr\Big\{\Big[\tilde\Lambda (\tilde F_0-zp\eta_gF_g)\Big]^{Rt/p}
\rho_f(0)\Big\}\Big|_{z=1}, \nonumber \\
\label{eq:eq3_14}
\end{eqnarray}
and
\begin{eqnarray}
\langle N_g(N_g-1)\rangle_t & = & {d^2\over dz^2}
Tr\Big\{\Big[\tilde\Lambda \Big(\tilde F_0-zp\eta_gF_g \Big)\Big]^{Rt/p}
\rho_f(0)\Big\}\Big|_{z=1}. \nonumber \\
\label{eq:eq3_15}
\end{eqnarray}
with similar expressions for $e$ rather than $g$. We use
(\ref{eq:eq3_12})--(\ref{eq:eq3_15}) to evaluate the parameters $Q_{g}$
defined in Mandel's \cite{r31} $Q$-parameter form by
\begin{eqnarray}
Q_{g} = {\langle N_{g} (N_{g} - 1) \rangle \over \langle N_{g} \rangle}
- \langle N_{g} \rangle.
\label{eq:eq3_16}
\end{eqnarray}
Definition (\ref{eq:eq3_16}) has a similar form for $Q_{e}$ with $e$
replacing $g$. These quantities (\ref{eq:eq3_16}) are measures of the
deviation of the observed distribution from a Poisson distribution. The
values $Q_{g} = 0$ means that the corresponding statistical distribution
is Poissonian; $Q_{g} < 0$ and $Q_{g} > 0$ imply sub- and super-Poissonian
distributions respectively. An approximate relationship between $Q_{g}$
and $Q_{f}$ the corresponding $Q$-function for the cavity field both
evaluated for fixed $t$ is derived in ref. \cite{r21}. However this
relation has since been found to be of limited validity \cite{r25,r26}.
The numerical calculations described in the next section for both fixed
$t$ and fixed $N$ lead us to a similar conclusion. 

Now, the operations of differentiation in
(\ref{eq:eq3_12})--(\ref{eq:eq3_15}) can be performed by using the easily
verifiable results that
\begin{eqnarray}
{d\over dx}(A+xB)^n & = & \sum_{k=0}^{n-1}(A+xB)^kB(A+xB)^{n-k-1}, \nonumber \\
\label{eq:eq3_17}
\end{eqnarray} 
and
\begin{eqnarray}
& & {d^2\over dx^2}(A+xB)^n = \sum_{k,l} [(A+xB)^lB(A+xB)^{k-l-1} \nonumber \\
& & \times B (A+xB)^{n-k-1} + (A+xB)^kB(A+xB)^l \nonumber \\
& & \times ~~~~~~B(A+xB)^{n-k-l-2}].
\label{eq:eq3_18}
\end{eqnarray}
In this paper we shall evaluate (\ref{eq:eq3_12})--(\ref{eq:eq3_15}) by
assuming, in accordance with ref. \cite{r26}, that the system is initially
in the steady state, i.e., $\rho_f(t_1)=\rho_f(0)=\rho_{ss}$ where
$\rho_{ss}$ is the solution of (\ref{eq:eq2_40}). We make use of the fact
that $D_p$ and $F_0$ are both trace conserving, i.e.,
\begin{eqnarray}
Tr_{f} [D_{p} \rho] = Tr_{f} [\rho] \mbox{  } \hbox{,} \mbox{  }
Tr_{f} [F_{0} \rho] = Tr_{f} [\rho],
\label{eq:eq3_19}
\end{eqnarray}
for an arbitrary $\rho$. On evaluating (\ref{eq:eq3_12}) and
(\ref{eq:eq3_13}) and substituting the resulting expressions in
(\ref{eq:eq3_16}) one finds
\begin{eqnarray}
\langle N_g\rangle_N=\eta_g N\hbox{ }Tr_f[F_g\rho_{ss}],
\label{eq:eq3_20}
\end{eqnarray} 
and
\begin{eqnarray}
Q_g & = & {2\eta_g\over Tr_f[F_g\rho_{ss}]}
Tr_f\Big[F_g\Big(1-{1-(\Lambda_pF_0)^N\over
N(1-\Lambda_pF_0)}\Big) \nonumber \\
& & \times \Big({1\over 
1-\Lambda_pF_0}\Big)\Lambda_pF_g\rho_{ss}\Big] 
- \eta_g N Tr_f[F_g\rho_{ss}],
\label{eq:eq3_21}
\end{eqnarray}
which is the desired expression for $Q_g$ with particular reference to the
detection of atoms in the ground state $|g\rangle$ for a fixed number $N$
of excited atoms entering the cavity in an arbitrary time. Similarly, in
the case when the observation is made for a fixed time $t$ the expressions
(\ref{eq:eq3_14}) and (\ref{eq:eq3_15}) yield
\begin{eqnarray}
\langle N_g\rangle_t=\eta_g Rt\hbox{ }Tr_f[F_g\rho_{ss}]
\label{eq:eq3_22}
\end{eqnarray} 
and
\begin{eqnarray}
\tilde Q_g&=&{2\eta_gp^2\over Tr_f[F_g\rho_{ss}]} Tr_f
\Big[F_g\Big({1\over p}-{1-(\tilde\Lambda_p\tilde F_0)^{Rt/p}
\over Rt(1-\tilde\Lambda_p\tilde F_0)}\Big) \nonumber \\
& & \times \Big({1\over 1-\tilde\Lambda_p\tilde F_0}\Big)\tilde\Lambda_pF_g
\rho_{ss}\Big]-\eta_g Rt Tr_f[F_g\rho_{ss}], \nonumber \\
\label{eq:eq3_23}
\end{eqnarray}
where the tilde on the $Q$ now means fixed collection time $t$.  In the
Poisson limit $p\rightarrow 0$, (\ref{eq:eq3_22}) and (\ref{eq:eq3_23})
reduce to the corresponding expressions derived for fixed $t$ in ref.
\cite{r26}. The expressions for $Q_e$ and $\tilde Q_e$ are obtained by
replacing the quantities with suffix $g$ by the corresponding ones with
suffix $e$ in (\ref{eq:eq3_21}) and (\ref{eq:eq3_23}) respectively. Note
that all of these $Q$-functions are directly proportional to the
corresponding detector efficiencies for any choice of $p$, thus solving
the problem raised, for example, in \cite{r21} and, for $p \rightarrow 0$
and fixed $t$ in agreement with \cite{r26}. 

We outline in the Appendix a method of actually evaluating these
$Q$-functions (\ref{eq:eq3_21}) and (\ref{eq:eq3_23}) for both of $e$ and
$g$. 

We have already shown in the last section (at (\ref{eq:eq2_40})) that the
state of the {\em field} observed after the passage of a large number $N$
of active atoms is the same as the one observed after waiting for a long
time. We now find the atomic $Q$-functions for a group of a fixed number
$N\rightarrow \infty$ of atoms using (\ref{eq:eq3_21}) and for a group of
atoms passing through the cavity in a fixed time $t\rightarrow\infty$ by
evaluating (\ref{eq:eq3_23}) in the limit $t\rightarrow\infty$. The
results of the numerical computations are presented in the next section. 

Here we compare the predictions of the two approaches in the case of an
analytically solvable example. This is the example of the micromaser
operating without any thermal photons ($\bar n=0$ in (\ref{eq:eq2_2})) and
having an interaction time such that $gt_{int}=\pi/\sqrt{2}$. In this case
the Fock state $|1\rangle$ is a trapping state \cite{r4}. Hence, if the
cavity field is initially in the state of vacuum then the state of the
micromaser is described by two-dimensional matrices and this makes the
problem analytically tractable. We find the $Q$-functions for the upper
and the lower states in the limit $t\rightarrow\infty$ to be given by the
formulae
\begin{eqnarray}
\tilde Q_e & = & {-\eta_e\over 1-d+p\beta_1d} \Big({2p^2\beta^2_1(1-d)
(\alpha_1+p\beta_1d-d)d\over (1-d+p\beta_1d)[\alpha_1(1-d) + p\beta_{0}d]} 
\nonumber \\
& & + p[\alpha_1(1-d)+p\beta_1d]\Big),
\label{eq:eq3_24}
\end{eqnarray}
and
\begin{eqnarray}
\tilde Q_{g} & = & {-\eta_g\over 1-d+p\beta_1d} 
\Big(\frac{2p\beta_1(1-p\beta_1) (1-d)d}{1-d + p \beta_1d} \nonumber \\
& & + p\beta_1(1-d)\Big),
\label{eq:eq3_25}
\end{eqnarray}
where
\begin{eqnarray}
d & = & \exp \Big(-\frac{p}{N_{ex}}\Big), \nonumber \\
\beta_1 & = & 1-\alpha_1 = \sin^{2} \Big(\frac{\pi}{\sqrt{2}}\Big), 
\nonumber \\ N_{ex} & = & \frac{R}{2\kappa}.
\label{eq:eq3_26}
\end{eqnarray}
The parameter $N_{ex}$ is the usual so-called pumping parameter mentioned
in the abstract which gives the average number of atoms pumped into the
cavity in one cavity damping time. Note from (\ref{eq:eq3_25}) that
$\tilde Q_g$ is always negative, i.e., sub-Poissonian whereas $\tilde Q_e$
is definitely sub-Poissonian if $\alpha_1+p\beta_1d-d> 0$. For the
particular case of Poisson pumping statistics the limit $p\rightarrow 0$
in (\ref{eq:eq3_24}) and (\ref{eq:eq3_25}) yields
\begin{eqnarray}
\tilde Q_e={2\eta_e\beta^3_1N^2_{ex}\over (N_{ex}\beta_1+\alpha_1)
(N_{ex}\beta_1+1)^2},
\label{eq:eq3_27}
\end{eqnarray}
and
\begin{eqnarray}
\tilde Q_g={-2\eta_e\beta_1N_{ex}\over (N_{ex}\beta_1+1)^2}.
\label{eq:eq3_28}
\end{eqnarray}
These expressions (\ref{eq:eq3_27}) and (\ref{eq:eq3_28}) for fixed $t
\rightarrow \infty$ and $p \rightarrow 0$ are exactly the same as the
corresponding ones derived in ref. \cite{r26}. Thus, in the case of
Poisson pumping, the atoms exiting the cavity in the upper state follow a
super-Poisson distribution whereas those emerging in the lower state have
a strictly sub-Poissonian distribution. 

In a similar way the $Q$-functions in the limit $N\rightarrow\infty$ are
found to be given by
\begin{eqnarray}
Q_e & = & {-\eta_e\over 1-d+p\beta_1d} \Big({2p\beta^2_1\alpha_1(1-d)^2d
\over (1-d+p\beta_1d)[\alpha_1(1-d)+p\beta_1d]} \nonumber \\
& & + \alpha_1(1-d) + p \beta_1d\Big).
\label{eq:eq3_29}
\end{eqnarray}
and
\begin{eqnarray}
Q_g & = & {-\eta_g\over 1-d+p\beta_1d} \Big({2p\beta_1\alpha_1(1-d)d
\over 1-d+p\beta_1d}+\beta_1(1-d)\Big). \nonumber \\
\label{eq:eq3_30}
\end{eqnarray}
Note that in this case, i.e., in the case of observations made on groups
of a fixed large number $N$ of atoms as $N \rightarrow \infty$, both $Q_e$
as well as $Q_g$ are sub-Poissonian. In the Poisson limit $p\rightarrow
0$, (\ref{eq:eq3_29}) and (\ref{eq:eq3_30}) reduce to
\begin{eqnarray}
Q_e & = & {-\eta_e\over \beta_1N_{ex}+1} \Big({2\beta^2_1\alpha_1N_{ex}
\over (\beta_1N_{ex}+1)(\beta_1N_{ex}+\alpha_1)} \nonumber \\
& & +\beta_1N_{ex}+\alpha_1\Big),
\label{eq:eq3_31}
\end{eqnarray}
and
\begin{eqnarray}
Q_g & = & {-\eta_g\over \beta_1N_{ex}+1} \Big({2\beta_1\alpha_1N_{ex}
\over (\beta_1N_{ex}+1)}+\beta_1\Big).
\label{eq:eq3_32}
\end{eqnarray}
The differences between the atomic statistics in the long time and the
large number of atoms limits are clearly reflected in these various
analytical results. 

It is instructive to consider the behavior of the $Q$-functions derived
above for the special cases of the micromaser with low and high rates of
pumping respectively. For the low rate of pumping $R$ (or equivalently
high rate $\kappa$ of field damping), $N_{ex}\rightarrow 0$. In this limit
$\langle N_g\rangle \rightarrow N\eta_g\beta_1$, $\langle
N_e\rangle\rightarrow N\eta_e (1-\beta_1)$, $Q_g\rightarrow
-\eta_g\beta_1$, $Q_e\rightarrow -\eta_e(1-\beta_1)$. These values of
$\langle N_g\rangle$ ($\langle N_e\rangle$) and $Q_g$ ($Q_e$ ) correspond
to a binomial process in which the probability of success is $\beta_{g}$
$([1 - \beta_{g}])$ (see (\ref{eq:eq4_1}) below). Hence it follows that
the statistics of the atomic populations exiting the cavity is binomial
irrespective of the pumping statistics if the observations are made on
groups of a fixed number, $N\rightarrow\infty$, of atoms. If, on the other
hand, the observations are made for fixed time, $t\rightarrow\infty$ then
the desired averages are obtained from the corresponding ones for groups
of fixed number $N$ of atoms by averaging over $N$ with the probability
that $N$ atoms exit the cavity in time $t$. In this case, for the present
example, $\langle N_g\rangle$ ($\langle N_e\rangle$) and $Q_g$ ($Q_e$) are
expected to be $Rt\eta_g\beta_1$ ($Rt\eta_e [1-\beta_1 ]$), and $-p\eta_g
\beta_1$ ($-p\eta_e [1-\beta_1]$) [see (\ref{eq:eq4_2})]. It is easy to
see that (\ref{eq:eq3_24}) and (\ref{eq:eq3_25}) indeed yield those
results in the limit $N_{ex}\rightarrow 0$, so that $d \rightarrow 0$. 

Next, for $N_{ex} \rightarrow \infty$, that is in the limit of a high rate
of pumping or equivalently of a low rate of field losses (namely for an
ideal cavity), one has $Q_{e} \rightarrow -1$, $\tilde Q_{e} \rightarrow
-p$, $Q_{g} \rightarrow \tilde Q_{g} \rightarrow 0$. These are evidently
precisely the values of the $Q$-parameters for the atoms entering the
cavity. This behavior of the $Q$-parameters can be readily understood by
noting that in this limit, the steady state of the field is the trapping
state $\vert 1\rangle$: the atoms therefore exit the cavity in their upper
states $\vert e\rangle$ which are the states in which they enter the
cavity. Hence, in the limit $N_{ex}\rightarrow \infty$, the statistics of
the atoms exiting the cavity is exactly the same as the pumping
statistics. 

In the next section we investigate in detail the differences between the
atomic statistics in the long time and in the large $N$ limit by numerical
evaluation of not only the analytical expressions for the $Q$-functions
for the special case of $gt_{int} = \pi/\sqrt{2}$ at zero temperature
discussed above, but also for several other $gt_{int}$'s defining both
trapping states and non-trapping states \cite{f4} at both zero and finite
temperatures. 

\section{numerical results and discussion}
\label{sec:sec4}

In this section we present numerical evaluations of the atomic $Q$
functions in the asymptotic limits $N\rightarrow\infty$ and
$t\rightarrow\infty$. Since the efficiency factors appear only as a
multiplicative factor we give the results only for unit efficiencies
$\eta_{e} = \eta_{g} = 1$. 

\begin{figure}[t]
\centerline{\epsfxsize=\columnwidth \epsfbox[40 150 550 590]{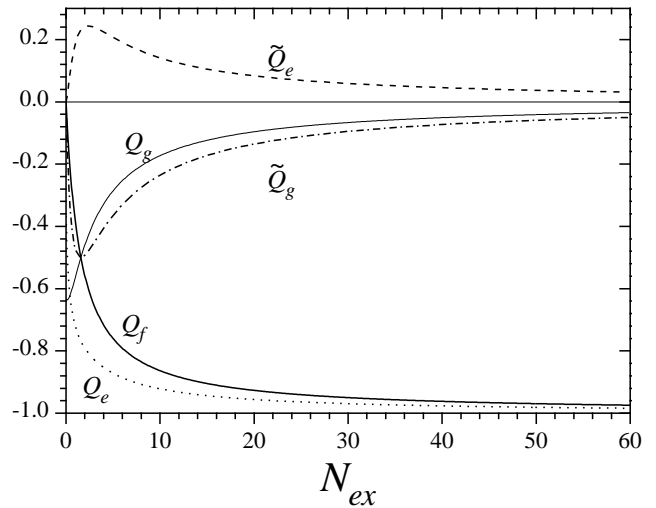}}
\caption{The $Q$-parameters, $Q_{g}$, $Q_{e}$, $\tilde Q_{g}$, $\tilde
Q_{e}$, and $Q_{f}$ as functions of $N_{ex}$ for $gt_{int} = \pi/
\protect\sqrt{2}$, $\bar{n} = 0$ and for Poissonian pumping of the
excited atoms into the cavity ($p=0$). $Q_{f}$ is multiplied by $0.1$
in order to compare it with the atomic $Q$-parameters.}
\label{fig:fig1}
\end{figure}

The parameter $Q_e$ ($Q_g$) is obviously the same as $\tilde Q_e$ ($\tilde
Q_g$) for the case of regular pumping $p=1$. We, therefore, examine these
parameters first for $p=0$ (Poisson pumping) for this corresponds to a
maximum departure from the regular input. We display first in Fig.
\ref{fig:fig1} the plots of $Q_{e}$, $Q_{g}$ and $\tilde Q_{e}$, $\tilde
Q_{g}$ together with $Q_{f}$ as functions of the pumping parameter
$N_{ex}=R/2\kappa$ in the case of a Poisson input and for which
$gt_{int}=\pi/\sqrt{2}$ while the system is at zero temperature, $\bar
n=0$. The exact analytical expressions for these $Q$-parameters were given
towards the end of the last section. It is clear from the figure that
$Q_e$ and $\tilde Q_e$ differ in the entire range of $N_{ex}$ whereas
$Q_g$ and $\tilde Q_g$ differ significantly for small $N_{ex}$ but
approach each other as $N_{ex}$ is increased. Notice how it is only
$Q_{e}$ which closely follows $Q_{f}$. This is because $Q_{g}$, $\tilde
Q_{e}$ and $\tilde Q_{g}$ all tend to zero with $N_{ex} \rightarrow
\infty$ while $Q_{e} \rightarrow -1$. Notice that as $N_{ex} \rightarrow
\infty$ either cavity $Q \rightarrow \infty (2\kappa \rightarrow 0)$ or
the mean entry rate $R \rightarrow \infty$ (or both). If $R \rightarrow
\infty$ the probability of there being more than one atom in the cavity
becomes large and the TC-model Hamiltonian \cite{r16} must replace the
JC-Hamiltonian Eq. (\ref{eq:eq2_1}). For $N_{ex} \rightarrow \infty$ we
must therefore assume the cavity $Q$ becomes large so as to maintain any
relevance to the actual experiments done so far. 

\begin{figure}[b]
\centerline{\epsfxsize=\columnwidth \epsfbox[40 150 550 590]{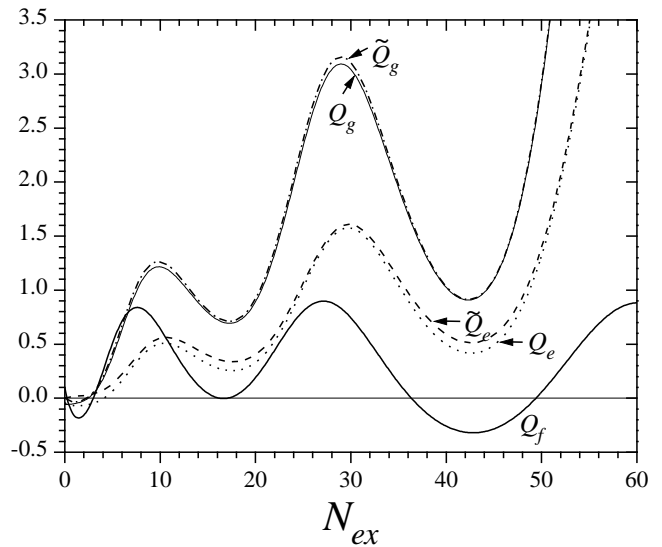}}
\caption{The same as Fig. \protect\ref{fig:fig1}, but for non-zero thermal
photons ($\bar{n} = 0.1$).}
\label{fig:fig2}
\end{figure}

We next examine the effect of finite temperatures on the differences in
the two kinds of $Q$-functions by plotting in Fig. \ref{fig:fig2} the same
data as in Fig. \ref{fig:fig1} but for $\bar n=0.1$. The differences
between the two kinds of $Q$-functions seem to decrease with an increase
in $\bar n$. The other point to notice is that $Q_{e}$ and $Q_{f}$ were
both sub-Poissonian for all $N_{ex}$ and were monotonically decreasing to
$-1$ at zero temperature as in the Fig. \ref{fig:fig1}. These behaviors
for both $Q_{e}$ and $Q_{f}$ and indeed the behaviors of $Q_{g}$, $\tilde
Q_{e}$ and $\tilde Q_{g}$ evidently change dramatically with the increase
in temperature $0.2$ K $(\bar{n} = 0.10)$. We return to this in the Figs.
\ref{fig:fig10}--\ref{fig:fig13} where these effects of temperature are
displayed. 

\begin{figure}[t]
\centerline{\epsfxsize=\columnwidth \epsfbox{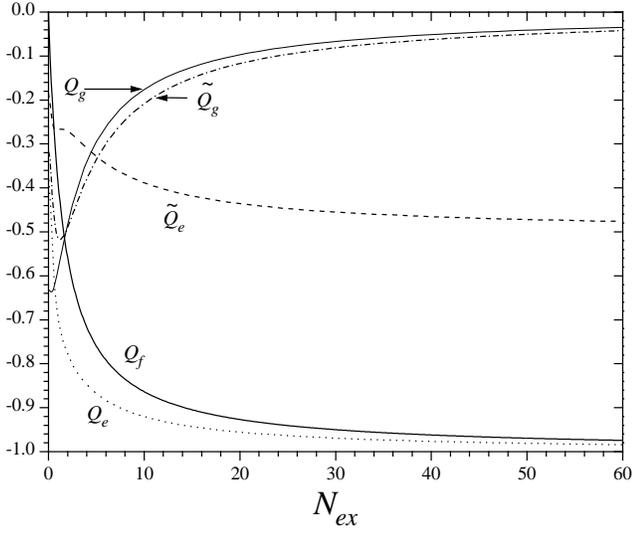}}
\caption{The same as Fig. \protect\ref{fig:fig1}, but for binomial pumping
of the excited atoms into the cavity with excitation probability, $p=0.5$
(deviation from Poissonian pumping).}
\label{fig:fig3}
\end{figure}

In Fig. \ref{fig:fig3} we investigate the effects of the departure from
the continuous time Poisson statistics on the differences between the two
limits by plotting the $Q$-functions as functions of $N_{ex}$ for the
discrete time binomial distribution with $p = 0.5$ for $gt_{int} =
\pi/\sqrt{2}$ and $\bar n = 0$. Not too much difference from the plots of
Fig. 1 can be seen from this significant change of $p$, except for the 
observation that $\tilde{Q}_{e}$ shows sub-Poissonian behavior for the 
entire range of $N_{ex}$, whereas for $p = 0$, it was totally 
super-Poissonian.

\begin{figure}[b]
\centerline{\epsfxsize=\columnwidth \epsfbox{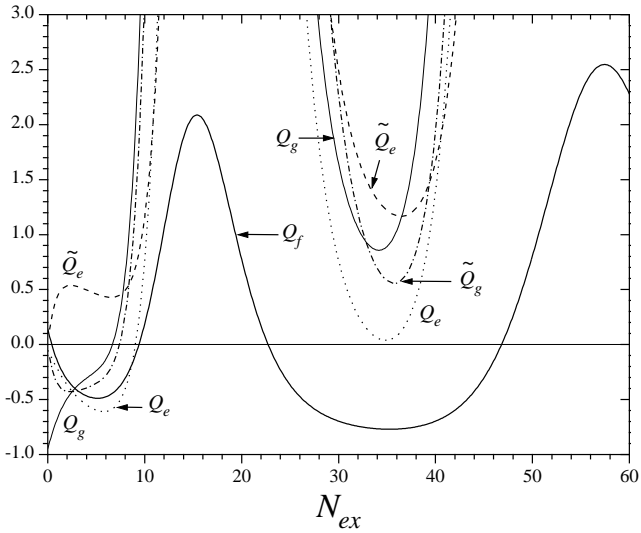}}
\caption{The $Q$-parameters, $Q_{g}$, $Q_{e}$, $\tilde Q_{g}$, $\tilde
Q_{e}$, and $Q_{f}$ as functions of $N_{ex}$ for $gt_{int} = 1.54$ and
$\bar{n} = 0.145$ and for Poisson pumping of the excited atoms into the
cavity ($p=0$). These parameters correspond to the Fig.
\protect\ref{fig:fig3} of \protect\cite{r20}.}
\label{fig:fig4}
\end{figure}

Next we plot in Fig. \ref{fig:fig4} the $Q$-functions for another value of
$gt_{int}$ namely, $gt_{int} = 1.54$ as a function of $N_{ex}$ for $\bar
n=0.145$ (temperature $ = 0.5$ K) and for the case of Poissonian pumping.
This is the set of parameters used in the experiments of ref. \cite{r20}.
Once again we find that the two kinds of $Q$-functions differ considerably
and differ considerably from $Q_{f}$ for small $N_{ex}$. It is interesting
to compare the plot of $\tilde Q_g$ with the experimental plot, Fig.
\ref{fig:fig3} of ref. \cite{r20}. One can check that the theoretical plot
for $\tilde Q_g$ given in Fig. \ref{fig:fig4} here exhibits the minima and
the maxima at the same positions as the plot of the experimental points.
However, whereas the second minima of the experimental points exhibits
sub-Poisson behavior, that of the theoretical plot is super-Poissonian and
to this extent the sub-Poissonian field with $Q_{f} = -0.3$ observed at
$N_{ex} = 35$ in $\tilde Q_{g}$ in \cite{r20} may need further scrutiny.
Note that the numerical values in the two plots, even after correcting for
the efficiency factor, do not match at all. The reason for the discrepancy
between the two apparently may stem from the fact that there are stray
fields present in the experiments (cf. \cite{r42}) which have not been
taken into account in the theoretical plot of Fig. \ref{fig:fig4}. 

\begin{figure}
\centerline{\epsfxsize=\columnwidth \epsfbox{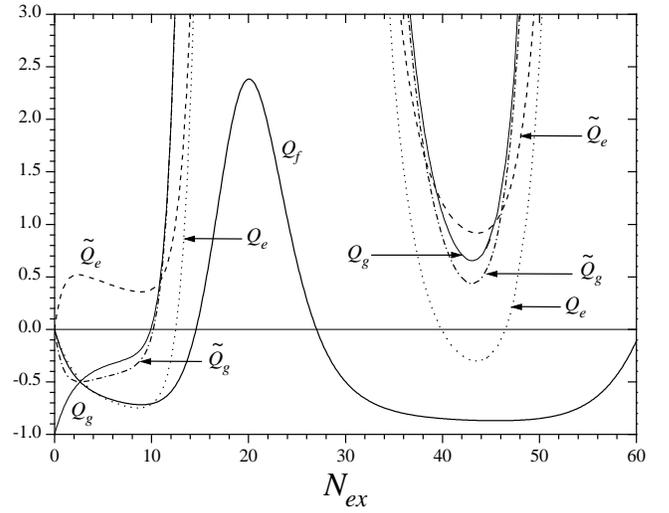}}
\caption{The same as Fig. \protect\ref{fig:fig4}, but for $\bar{n} = 0$.}
\label{fig:fig5}
\end{figure}

In Fig. \ref{fig:fig5} we plot the same data as in Fig. \ref{fig:fig4}
except that $\bar{n} = 0$ (zero temperature). In view of the Figs.
\ref{fig:fig1} and \ref{fig:fig2} where $\bar{n} = 0$ and $\bar{n} > 0$
the most striking feature of Fig. \ref{fig:fig5} ($\bar{n} = 0$) in
relation to Fig. \ref{fig:fig4} ($\bar{n} > 0$) is that qualitatively
there is little change, although $Q_{e}$ for $N_{ex} \sim 44$ is
sub-Poissonian and approaching $Q_{f}$ from Fig. \ref{fig:fig4} which is
also sub-Poissonian for however $N_{ex} \approx 35$. Note that $gt_{int} =
1.54$ is not a low-lying (in terms of photon number $n$) trapping state. 

Notice how the increase in temperature from zero to $0.15$ K between the
Figs. \ref{fig:fig5} and \ref{fig:fig4} pushes all of the $Q$'s
substantially to the left (lower $N_{ex}$) so that each of the minima of
$Q_{e}$, $\tilde Q_{e}$, $Q_{g}$, $\tilde Q_{g}$, of which only $Q_{e}$ is
sub-Poissonian, arise at $N_{ex} \sim 44$ in Fig. \ref{fig:fig5} while
these are pushed to $N_{ex} \sim 35$ in Fig. \ref{fig:fig4} with now
$Q_{e}$ just super-Poissonian. Apparently the big maximum for the atomic
$Q$'s for $N_{ex} \ge 42$ in Fig. \ref{fig:fig4} is pushing these minima
strongly to the left as described. Notice now that this effect of
temperature for $gt_{int} = 1.54$ is {\em qualitatively} different from
the effect of temperature between Fig. \ref{fig:fig1} and Fig.
\ref{fig:fig2}. This is apparently due to the fact that in Figs.
\ref{fig:fig1} and \ref{fig:fig2}, $gt_{int} = \pi / \sqrt{2}$ which is
the one-photon trapping state condition: the $gt_{int} = 1.54$ radians
does not define any trapping state with a small number of photons. Thus,
with $\pi = 3.14159$ for example, one finds that the lowest trapping state
has the photon number $n \sim 4 \times 10^{10}$! At zero temperature and
$T_{c} = \infty$ (ideal cavity) the trapping states are the only
equilibrium states \cite{r13,r19}; but although $gt_{int} = 1.54$ at zero
temperature in Fig. \ref{fig:fig5}, $T_{c} < \infty$ and equilibrium is
expected to lie below the {\em lowest} trapping state (in terms of photon
number) as was demonstrated in \cite{r13}. For lowest trapping states with
small photon number switching on the temperature induces a rather
remarkable `escape' to the larger photon numbers which leads to the rather
dramatic rises in variances evident in the sequence of thresholds for
these variances shown in the contrasts between the Figs. \ref{fig:fig5}
and shown below for which figures all are taken at low lying trapping
states. A conclusion remains that there is no simple relation between
$Q_{f}$ and any of $Q_{g}$, $Q_{e}$, $\tilde Q_{g}$ and $\tilde Q_{e}$ at
finite temperature while it is only $Q_{e}$ which follows $Q_{f}$
relatively closely at zero temperature. 

\begin{figure}
\centerline{\epsfxsize=\columnwidth \epsfbox{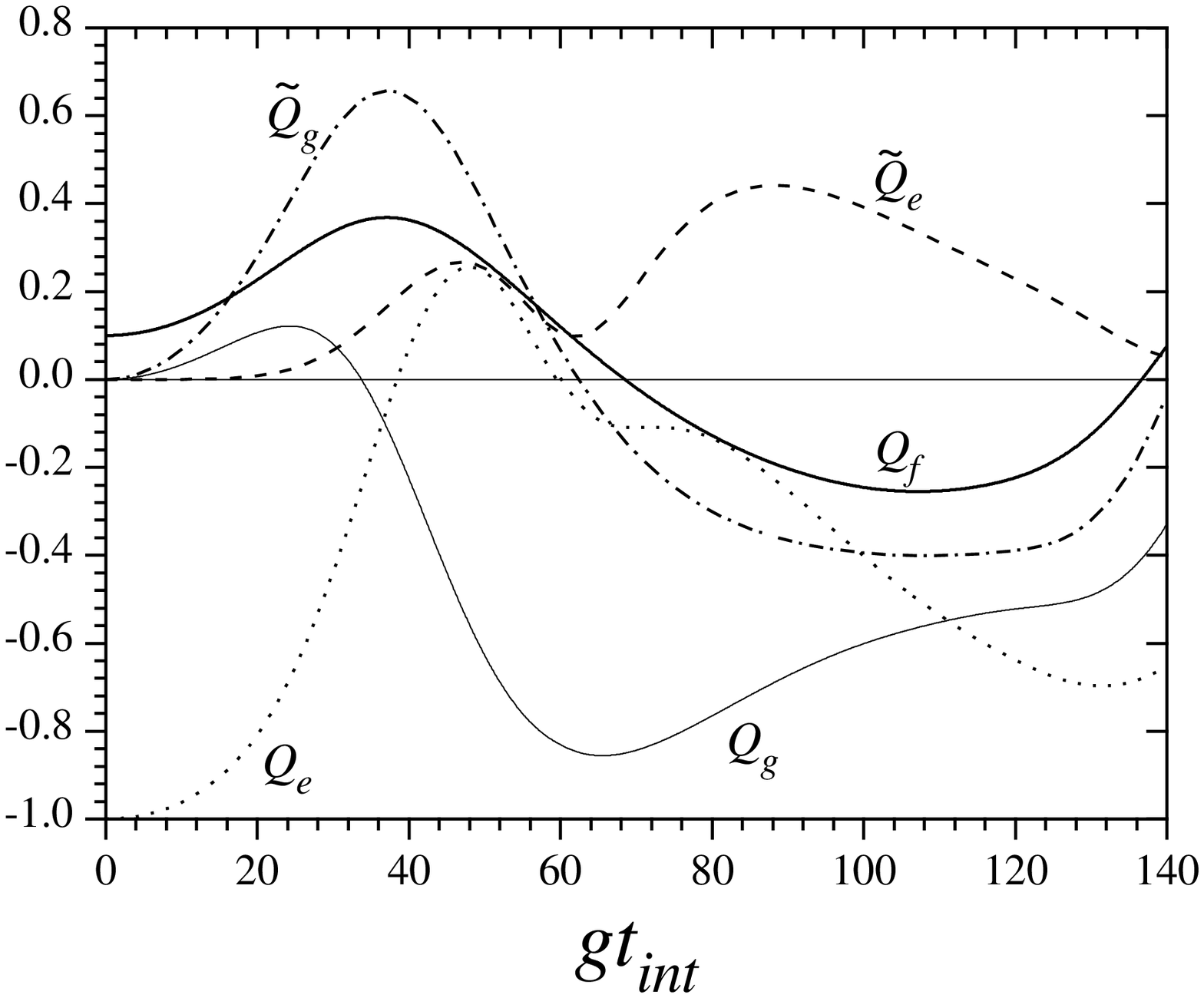}}
\caption{The $Q$-parameters, $Q_{g}$, $Q_{e}$, $\tilde Q_{g}$, $\tilde
Q_{e}$, and $Q_{f}$ as functions of $gt_{int}$ when $\bar{n} = 0.1$
and $N_{ex}=1$ for the case of Poissonian pumping of the excited
atoms.}
\label{fig:fig6}
\end{figure}

\begin{figure}
\centerline{\epsfxsize=\columnwidth \epsfbox{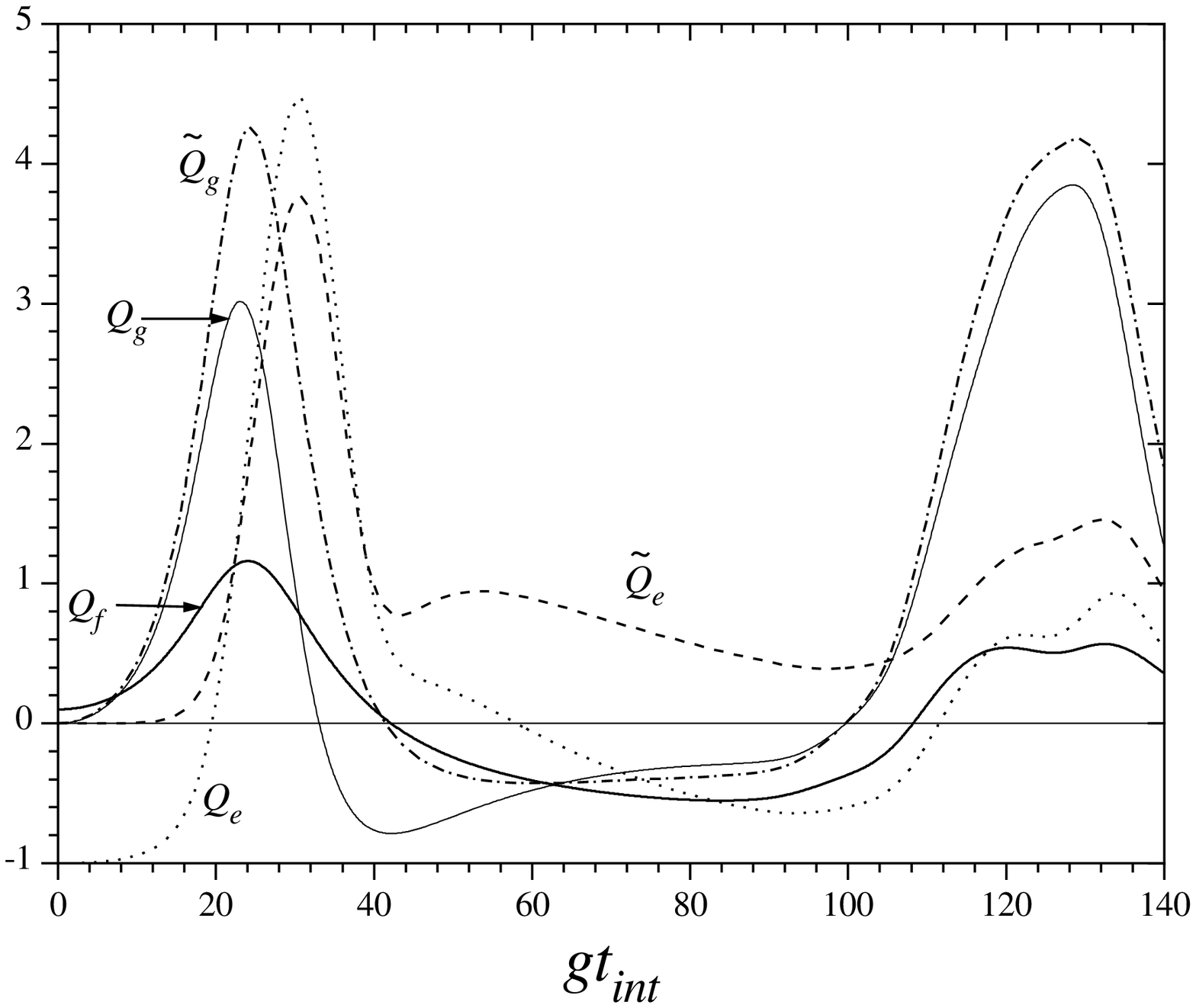}}
\caption{The $Q$-parameters, $Q_{g}$, $Q_{e}$, $\tilde Q_{g}$, $\tilde
Q_{e}$, and $Q_{f}$ as functions of $gt_{int}$ when $\bar{n} = 0.1$
and $N_{ex}=5$ for the case of Poissonian pumping of the excited
atoms.}
\label{fig:fig7}
\end{figure}

\begin{figure}
\centerline{\epsfxsize=\columnwidth \epsfbox{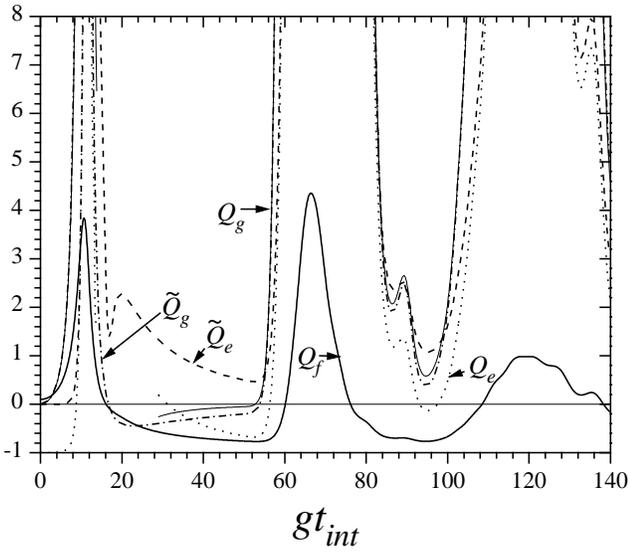}}
\caption{The same as Figs. \protect\ref{fig:fig6} and
\protect\ref{fig:fig7}, but for $N_{ex} = 30$. Computational difficulties
have prevented us following $Q_{e}$ for $17 \protect\alt gt_{int} 
\protect\alt 30$.}
\label{fig:fig8} 
\end{figure}

\begin{figure}
\centerline{\epsfxsize=\columnwidth \epsfbox{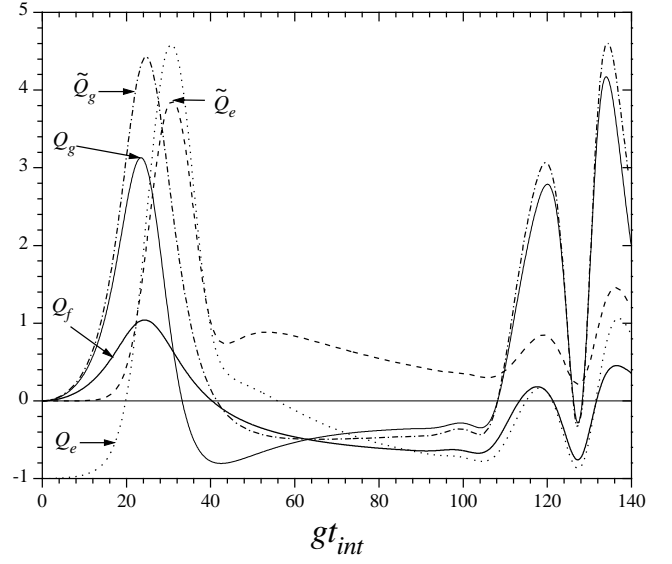}}
\caption{The same as Fig. \protect\ref{fig:fig7} but with $\bar{n} = 0$.}
\label{fig:fig9}
\end{figure}

\begin{figure}
\centerline{\epsfxsize=\columnwidth \epsfbox{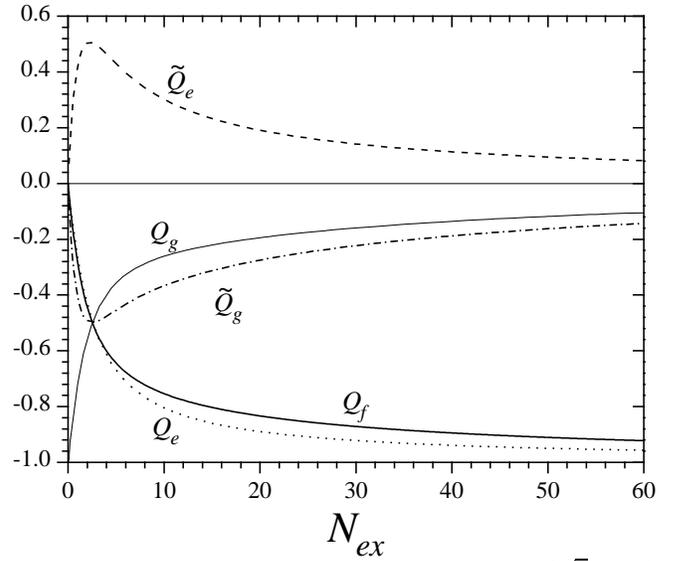}}
\caption{The same as Fig. \protect\ref{fig:fig1} but for $gt_{int} = $
$\pi/\protect\sqrt{4}$ $= \pi/2$ and $\bar{n} = 0$.}
\label{fig:fig10}
\end{figure}

\begin{figure}
\centerline{\epsfxsize=\columnwidth \epsfbox{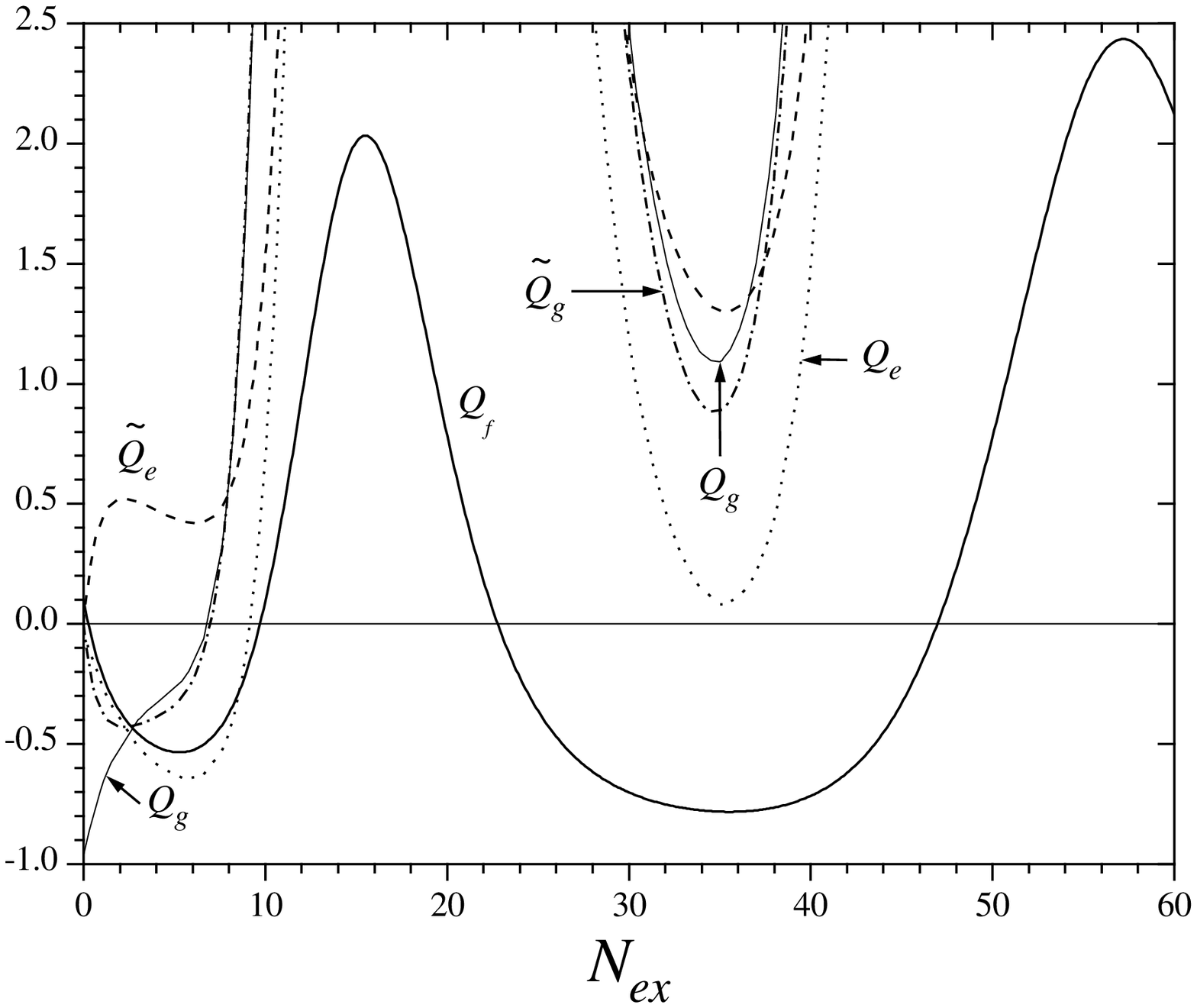}}
\caption{The same as Fig. \protect\ref{fig:fig10} but for finite
temperature ($\bar{n} = 0.1$).}
\label{fig:fig11}
\end{figure}

\begin{figure}
\centerline{\epsfxsize=\columnwidth \epsfbox{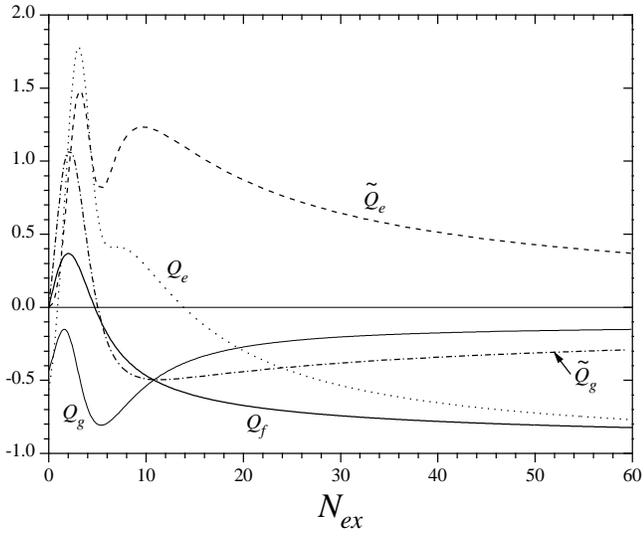}}
\caption{The same as Fig. \protect\ref{fig:fig1} but for $gt_{int} =
\pi/\protect\sqrt{19}$ and $\bar{n} = 0$.}
\label{fig:fig12}
\end{figure}

\begin{figure}
\centerline{\epsfxsize=\columnwidth \epsfbox{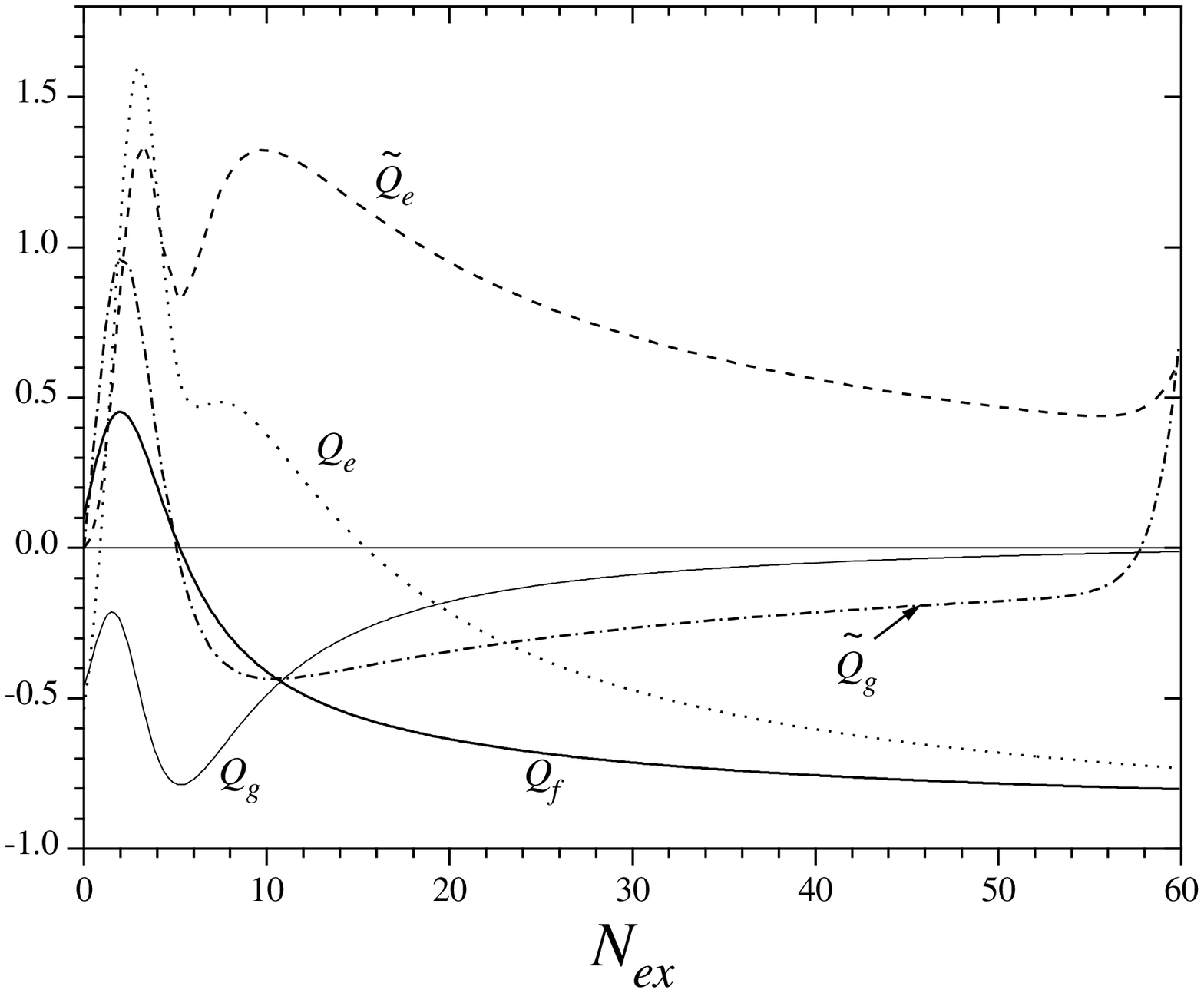}}
\caption{The same as Fig. \protect\ref{fig:fig12} but for finite
temperature ($\bar{n} = 0.1$).}
\label{fig:fig13}
\end{figure}

We next examine, in Figs. \ref{fig:fig6} and \ref{fig:fig7}, the
differences in the two kinds of asymptotic limits as a function of
$gt_{int}$ for $N_{ex}=1$ (Fig. \ref{fig:fig6}) and $N_{ex} = 5$ (Fig.
\ref{fig:fig7}) for $\bar n=0.1$ and for the case of Poisson pumping.
These figures confirm that these differences depend on $N_{ex}$ for any
$gt_{int}$. Thus the curves for $Q_{e}$, $Q_{g}$ differ from the
corresponding $\tilde Q_{e}$, $\tilde Q_{g}$ curves for all $gt_{int}$ for
$N_{ex}=1$ (Fig. \ref{fig:fig6}), while their differences decrease as the
values of $N_{ex}$ is increased to 5 (Fig. \ref{fig:fig7}). The Fig.
\ref{fig:fig8} shows these differences for the experimentally realized
value of $N_{ex} = 30$.  In the Fig. \ref{fig:fig9} we repeat the data of
Fig. \ref{fig:fig7} at $\bar{n} = 0$ (zero temperature). Low-lying
trapping states arise at $gt_{int} = \pi/\sqrt{2}$, $\pi/\sqrt{3}$,
$\pi/2$, $\pi/\sqrt{5}$, $\pi/\sqrt{6}$ for example for $r = 1$ in the
trapping state condition $gt_{int} \sqrt{n + 1} = r \pi$. For $r = 1, 2,
3$ etc. they arise for $gt_{int} = \pi/2$, $\pi/4$, $\pi/6$ and so on, for
example. Still Fig. \ref{fig:fig7} shows little qualitative change from
Fig. \ref{fig:fig9} ($\bar{n} = 0$). 

For the Fig. \ref{fig:fig10} we note that the value of $gt_{int}$,
$gt_{int} = \pi/\sqrt{2}$, in the Fig. \ref{fig:fig1} is a small value for
a trapping state. In the Fig. \ref{fig:fig10} we choose $\bar{n} = 0$
(zero temperature) and $gt_{int} = \pi/2$ radians. This is a trapping
state with photon number $n = 3$ also a small value. The behaviors in Fig.
\ref{fig:fig10} are qualitatively the same as in the Fig. \ref{fig:fig1}.
In the Fig. \ref{fig:fig11} the temperature is increased so that $\bar{n}
= 0.1$. The qualitative change compared with Fig. \ref{fig:fig10} follows
the qualitative change between Fig. \ref{fig:fig1} (zero temperature) and
Fig. \ref{fig:fig2} (finite temperature). 

For the Fig. \ref{fig:fig12} we choose $gt_{int} = \pi/\sqrt{19}$ so that
the trapping state photon number is increased to $n = 18$, larger than in
Figs. \ref{fig:fig1} and \ref{fig:fig2} and \ref{fig:fig10} and
\ref{fig:fig11}. The Fig. \ref{fig:fig13} shows the same data for $\bar{n}
= 0.1$; the change here is the emergence of the threshold for $\tilde
Q_{e}$, $\tilde Q_{g}$ just beyond $N_{ex} = 60$. 

\begin{figure}
\centerline{\epsfxsize=\columnwidth \epsfbox{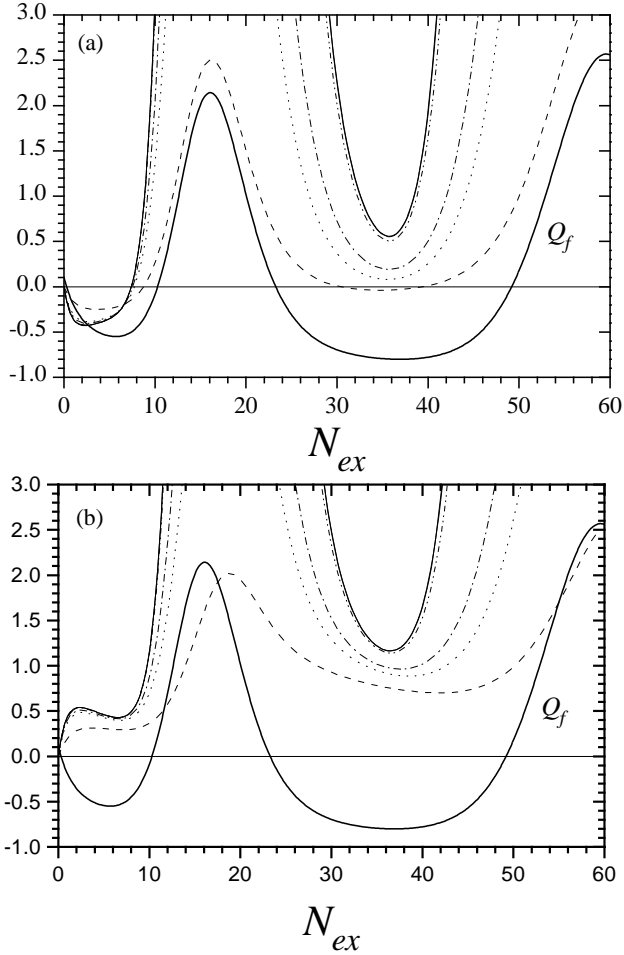}}
\caption{The dependence of (a) $\tilde Q_{g}$ and (b) $\tilde Q_{e}$ on
the collection time $t$ for $gt_{int} = 1.54$ and $\bar{n} = 0.1$. The
$t$ values are 1 ($- - -$), 5 ($\cdots$), 10 ($- \cdot -$), 100 ($- \cdot 
\cdot -$), 1000 (- - -), and $\infty$ (---).} 
\label{fig:fig14}
\end{figure}

\begin{figure}
\centerline{\epsfxsize=\columnwidth \epsfbox{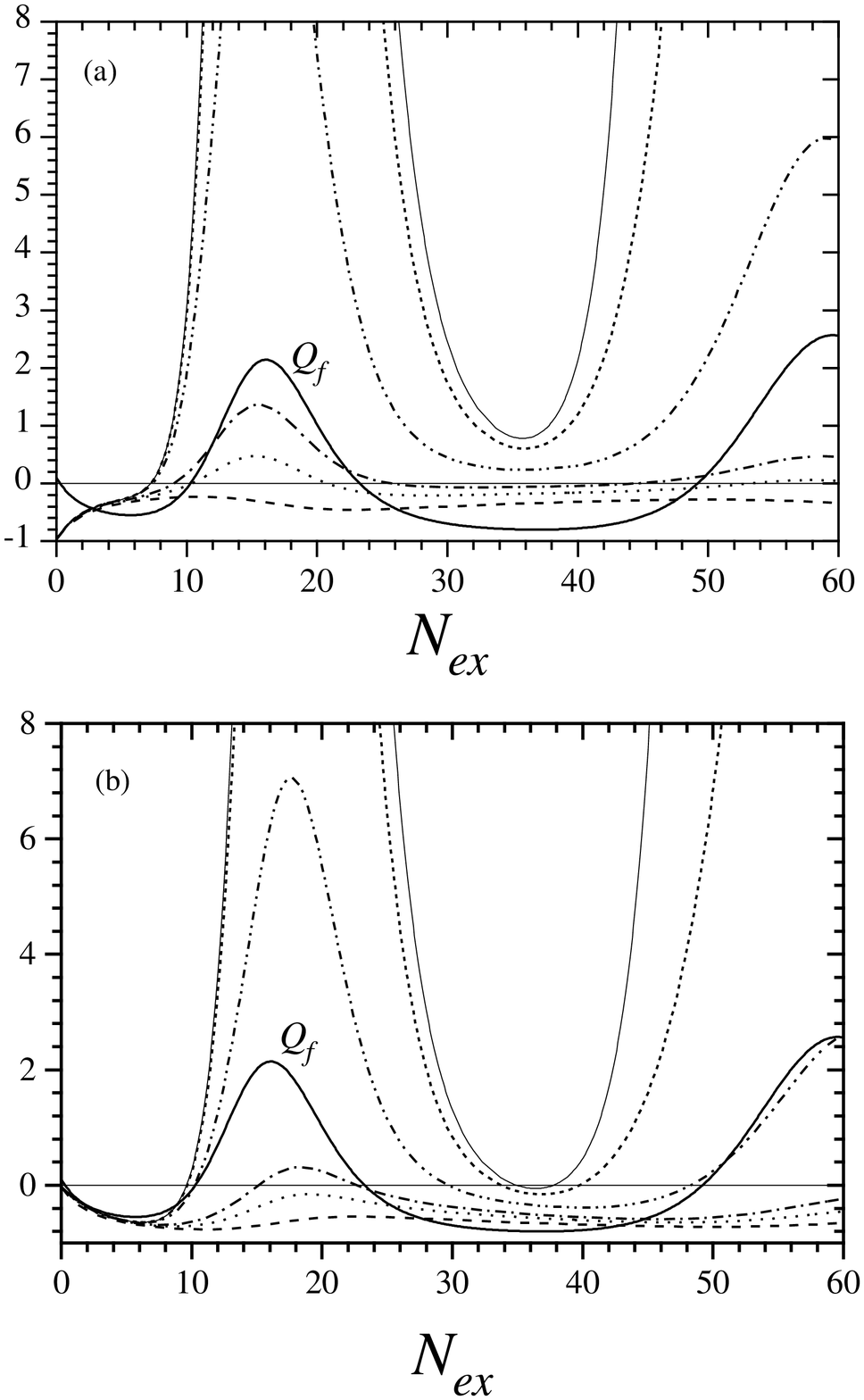}}
\caption{The dependence of (a) $Q_{g}$ and (b) $Q_{e}$ on the number of
atoms collected, $N$ for $gt_{int} = 1.54$ and $\bar{n} = 0.1$. The $N$
values are 20 ($- - -$), 50 ($\cdots$), 100 ($- \cdot -$), 500 ($- \cdot 
\cdot -$), 1000 (- - -), and $\infty$ (---).}
\label{fig:fig15}
\end{figure}

The Figs. \ref{fig:fig14} (a) and (b) plot the dependence of $\tilde
Q_{g}$ and $\tilde Q_{e}$ respectively on the collection time $t$ for
$gt_{int} = \pi/\sqrt{2}$. The chosen $t$ values are 1, 5, 10, 100, 1000
and $\infty$ times the cavity damping time and the plots are given as
functions of $N_{ex}$. These results compare with the Fig. \ref{fig:fig5}
in \cite{r42} evaluated for $gt_{int} = 1.54$ and plotted as a function of
collection time $t$. Finally the Figs. \ref{fig:fig15} (a) and (b) plot
the dependence of $Q_{g}$ and $Q_{e}$ respectively on $N$ the fixed number
of atoms collected as a function of $N$ for $gt_{int} = 1.54$ and $\bar{n}
= 0.1$. The chosen $N$ values are 20, 50, 100, 500, 1000, and $\infty$. 

The reason for the differences in the atomic $Q$-functions in the limit of
a large number of atoms and that in the limit of long time for $N_{ex} \ll
1$ can be easily seen. These limits correspond to high cavity damping or a
low rate of injection of the atoms. In either case the cavity field decays
to a state of thermal equilibrium during the time interval between the
arrival of two successive atoms into the cavity. Hence each atom sees the
same field, i.e. the thermal field, on its entry into the cavity. The
probability of an atom exiting the cavity in a given state is, therefore,
the same for every atom. In other words, there is no correlation between
the states of two atoms exiting from the cavity and so the probability
that an atom exits in a given state follows a binomial distribution. Let
$\beta_g$ be the probability that an atom exits from the cavity in its
ground state. If one observes groups of a fixed number $N$ of atoms
exiting from the cavity then the number $N_g$ of atoms detected in the
ground state is governed by a binomial distribution for which the
probability of success is $\beta_g$. Hence
\begin{eqnarray}
\langle N_g\rangle=N\beta_g, \qquad \langle N_g(N_g-1) \rangle =
N(N-1)\beta^2_g.
\label{eq:eq4_1}
\end{eqnarray}
As a consequence of (\ref{eq:eq4_1}) it follows that $Q_g=-\beta_g$. The
same relations hold for the $Q$-function $Q_e$ for the upper state with
$\beta_g$ replaced by $1-\beta_g$. For the exactly solvable case of the
micromaser discussed above, $\beta_g = \sin^2(\pi/\sqrt{2}) \approx 0.36$.
Hence $Q_g$ and $Q_e$ in this case are predicted to be approximately $-
0.64$ and $- 0.36$ respectively for $N_{ex}$ close to zero. This is in
agreement with the plots presented in Figs. \ref{fig:fig1}. 

If, on the other hand, one observes the groups of atoms for a fixed time
interval $t$ then the number $N$ of atoms entering the cavity in that
interval is governed by the pumping statistics. The expressions
corresponding to (\ref{eq:eq4_1}) in this case are obtained by averaging
(\ref{eq:eq4_1}) over the number $N$ of atoms. Now, the pumping statistics
has been assumed to be a binomial distribution for which the probability
of success is $p$. Hence $\langle N\rangle=Rt$ and $\langle
N^2\rangle-\langle N\rangle^2 =Rt(Rt-p)$. On averaging (\ref{eq:eq4_1})
over $N$ we get
\begin{eqnarray}
\langle N_g\rangle=Rt\beta_g, \qquad \langle N_g(N_g-1)\rangle =
Rt(Rt-p)\beta^2_g.
\label{eq:eq4_2}
\end{eqnarray}
As a result, $\tilde Q_g = -p\beta_{g}$. The same arguments hold for
$\tilde Q_e$ with $\beta_g$ replaced by $1-\beta_g$. In particular, for
the Poisson pumping $p=0$, it is expected that $\tilde Q_g=\tilde Q_e=0$
for $N_{ex}$ close to zero. This is clearly in agreement with the plots
presented above. 

As $N_{ex}$ increases, the correlations between the successive atoms start
dominating. In particular for $N_{ex} \rightarrow \infty$, i.e., in the
limit of an ideal cavity, the steady state of the cavity field approaches
a trapping state and all the atoms exit in the excited state which is the
state in which they enter the cavity. Hence, in this case, it is expected
that $Q_e\rightarrow -1$, $\tilde Q_e \rightarrow -p$, $Q_g\rightarrow
\tilde Q_g\rightarrow 0$. With the exception of $\tilde Q_{e}$ whose
asymptotic, large $N_{ex}$ behavior could not actually be shown in the
figures this behavior is borne out by the figures presented above. 

\acknowledgements

RRP gratefully acknowledges enlightening discussions with
Dr.~B.-G.~Englert and Prof. H.~Walther. Both RRP and SAK are grateful to
the UK's EPSRC for the financial support which has enabled this work to be
carried out. 

\appendix
\section{}

In this Appendix we outline a method for evaluating the $Q$-functions
given by (\ref{eq:eq3_21}) and (\ref{eq:eq3_23}). Note that those two
$Q$-parameters can be written as
\begin{eqnarray}
Q & = & {2\eta_g\over Tr_f[F_g\rho_{ss}]} Tr_f \Big[F_g \Big(1 -
{1 - (AB)^K \over K(1-AB)}\Big) \nonumber \\
& & \times \Big({1 \over (1 - AB)}\Big) AF_g\rho_{ss}\Big] - \eta_g K Tr_f
[F_g\rho_{ss}], 
\label{eq:eq_ap1}
\end{eqnarray}
where $K = N$, $A=\Lambda_p$, $B=F_0$ for $Q=Q_g$ and $K = Rt$,
$A=\tilde\Lambda_p$, $B=\tilde F_0$ for $\tilde Q=\tilde Q_g$. 

To evaluate (\ref{eq:eq_ap1}), it is convenient to work in the
eigen-representation of the operator $AB$. Let $\{\rho_i\}$ and
$\{\tilde\rho_i\}$ be the right and left eigenfunctions of $AB$
corresponding to the eigenvalues $\{\lambda_i\}$, i.e.,
\begin{eqnarray}
AB\rho_i & = & \lambda_i\rho_i, \qquad
\tilde\rho_iAB=\lambda_i\tilde\rho_i.
\label{eq:eq_ap2}
\end{eqnarray}
We assume that the eigenvalue $\lambda_0=1$, i.e., $\rho_0$ is the steady
state density matrix of the micromaser ($\rho_{ss}=\rho_0$). 

On taking the trace on both sides of the first of the equations in
(\ref{eq:eq_ap2}) and on using the trace conserving property of $A$ and
$B$, i.e., the relations $Tr(A\rho)=Tr(\rho)$, $Tr(B\rho)=Tr(\rho)$, which
are a consequence of the identification of $A$ and $B$ with the operators
$\Lambda_p$, $\tilde\Lambda_p$, $F_0$, $\tilde F_0$ and the trace
conserving property (\ref{eq:eq3_19}), it follows that
$Tr(\rho_i)=\lambda_iTr(\rho_i)$, and since $\lambda_0$ is the only
eigenvalue assumed to be unity, this leads to the result
\begin{equation}
Tr(\rho_i) = 0, \quad \hbox{if} \quad i\not=0.
\label{eq:eq_ap3}
\end{equation}
Now, the eigenstates $\{\rho_i\}$ and $\{\tilde\rho_i\}$ are orthogonal
with respect to the trace as a scalar product. Let us assume that they are
orthonormalized so that
\begin{equation}
Tr\Big(\tilde\rho_i\rho_j\Big) = \delta_{ij}.
\label{eq:eq_ap4}
\end{equation}
As a consequence of (\ref{eq:eq_ap4}), we can write
\begin{equation}
AB(\rho_i) = \sum_{i}C^g_{ij}(\rho_j),
\label{eq:eq_ap5}
\end{equation}
where
\begin{equation}
C^g_{ij} = Tr\Big(\tilde\rho_jAB\rho_i\Big).
\label{eq:eq_ap6}
\end{equation}
Now we can express the averages and the $Q$-functions in terms of the
eigenstates of $AB$ using (\ref{eq:eq_ap6}). 

Since $Tr(AF_g\rho) = Tr(F_g\rho)$, it follows on using (\ref{eq:eq_ap5}) 
that
\begin{equation}
Tr(F_g\rho) = C^g_{00}.
\label{eq:eq_ap7}
\end{equation}
In the same way, it can be shown that the expression (\ref{eq:eq_ap1}) for
$Q$ reduces to
\begin{eqnarray}
Q & = & -p \eta_{g} C^g_{00} + {2 \eta_{g}\over C^g_{00} K}
\sum_{i\not=0}\Big[{C^g_{i0}C^g_{0i}\over (1-\lambda_i)^2} 
\Big((1-\lambda_i) K-1 \nonumber \\
& & + [1-p(1-\lambda_i)]^K \Big)\Big].
\label{eq:eq_ap8}
\end{eqnarray}
The $Q$-functions can thus be computed by evaluating the right and the
left eigenstates and eigenvalues of the operator $AB$, where $A$ and $B$
are defined after (\ref{eq:eq_ap1}).

\vskip1.35in

\vskip1.35in

\vskip1.35in

\vskip1.35in

\vskip1.35in

\vskip1.35in

\newpage

\begin{references}
\bibitem{r30} We are indebted to B.-G.~Englert for first pointing out to
us the exact equivalence between `coarse-graining' and the choice of a
continuous time Poisson distribution for the pumping of the atoms. We
confirm this exact equivalence at both fixed $N$ and fixed $t$ in this
paper.
\bibitem{r38} It is wholly natural, and manipulatively convenient in
the present context as well, to introduce just such a concept, the
`conditional density matrix' (or `conditional density operator'). The
properties of this object are naturally defined by analogy with the
corresponding probability density (probability measure), and Eq.
(\ref{eq:eq2_11}), (\ref{eq:eq2_12}), (\ref{eq:eq2_13}) in particular
are examples of such natural definitions. The steps at
(\ref{eq:eq2_16}) and (\ref{eq:eq2_17}) to (\ref{eq:eq2_18}) and then
(\ref{eq:eq2_19}), with especially (\ref{eq:eq2_17}) defining joint and
conditional probabilities in their usual way, shows how natural this
idea of conditional density operators actually is. Evidently
(\ref{eq:eq2_19}) is finally a correct result, and the other
definitions (of conditional density operators) invoked serve only to
reach this result in a manipulatively convenient way. We believe these
ideas have been invoked by a number of different workers but cannot
give an explicit reference.
\bibitem{39} Recall $t_{i+1} - t_{i}$ is stochastic in practice: in
\cite{r20} for example, $R t_{int} \sim 1/200$.
\bibitem{f4} On the computer each value of $gt_{int}$ which is
achievable, namely a finite decimal number, defines a trapping state.
But such trapping states may require photon numbers $n$ very different
from the $n$'s considered.
\end{references}
\end{document}